\documentclass[twoside,twocolumn,9pt]{article}
\usepackage{extsizes}
\usepackage[super,sort&compress,comma]{natbib} 
\usepackage[version=3]{mhchem}
\usepackage[left=1.5cm, right=1.5cm, top=1.785cm, bottom=2.0cm]{geometry}
\usepackage{balance}
\usepackage{mathptmx}
\usepackage{sectsty}
\usepackage{graphicx} 
\usepackage{lastpage}
\usepackage[format=plain,justification=justified,singlelinecheck=false,font={stretch=1.125,small,sf},labelfont=bf,labelsep=space]{caption}
\usepackage{float}
\usepackage{fancyhdr}
\usepackage{fnpos}
\usepackage[english]{babel}
\usepackage{amsmath,amssymb}
\addto{\captionsenglish}{%
  \renewcommand{\refname}{Notes and references}
}
\usepackage{array}
\usepackage{droidsans}
\usepackage{charter}
\usepackage[T1]{fontenc}
\usepackage[usenames,dvipsnames]{xcolor}
\usepackage{setspace}
\usepackage[compact]{titlesec}
\usepackage{hyperref}

\newcommand{\rv}[1]{{\textcolor{black}{#1}}} 

\definecolor{cream}{RGB}{222,217,201}

\begin{document}

\pagestyle{fancy}
\thispagestyle{plain}
\fancypagestyle{plain}{
\renewcommand{\headrulewidth}{0pt}
}

\makeFNbottom
\makeatletter
\renewcommand\LARGE{\@setfontsize\LARGE{15pt}{17}}
\renewcommand\Large{\@setfontsize\Large{12pt}{14}}
\renewcommand\large{\@setfontsize\large{10pt}{12}}
\renewcommand\footnotesize{\@setfontsize\footnotesize{7pt}{10}}
\makeatother

\renewcommand{\thefootnote}{\fnsymbol{footnote}}
\renewcommand\footnoterule{\vspace*{1pt}%
\color{cream}\hrule width 3.5in height 0.4pt \color{black}\vspace*{5pt}} 
\setcounter{secnumdepth}{5}

\makeatletter 
\renewcommand\@biblabel[1]{#1}            
\renewcommand\@makefntext[1]%
{\noindent\makebox[0pt][r]{\@thefnmark\,}#1}
\makeatother 
\renewcommand{\figurename}{\small{Fig.}~}
\sectionfont{\sffamily\Large}
\subsectionfont{\normalsize}
\subsubsectionfont{\bf}
\setstretch{1.125} 
\setlength{\skip\footins}{0.8cm}
\setlength{\footnotesep}{0.25cm}
\setlength{\jot}{10pt}
\titlespacing*{\section}{0pt}{4pt}{4pt}
\titlespacing*{\subsection}{0pt}{15pt}{1pt}

\fancyfoot{}
\fancyfoot[LO,RE]{\vspace{-7.1pt}\includegraphics[height=9pt]{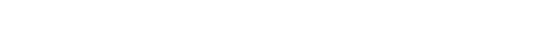}}
\fancyfoot[CO]{\vspace{-7.1pt}\hspace{13.2cm}\includegraphics{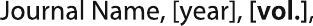}}
\fancyfoot[CE]{\vspace{-7.2pt}\hspace{-14.2cm}\includegraphics{head_foot/RF}}
\fancyfoot[RO]{\footnotesize{\sffamily{1--\pageref{LastPage} ~\textbar  \hspace{2pt}\thepage}}}
\fancyfoot[LE]{\footnotesize{\sffamily{\thepage~\textbar\hspace{3.45cm} 1--\pageref{LastPage}}}}
\fancyhead{}
\renewcommand{\headrulewidth}{0pt} 
\renewcommand{\footrulewidth}{0pt}
\setlength{\arrayrulewidth}{1pt}
\setlength{\columnsep}{6.5mm}
\setlength\bibsep{1pt}

\makeatletter 
\newlength{\figrulesep} 
\setlength{\figrulesep}{0.5\textfloatsep} 

\newcommand{\topfigrule}{\vspace*{-1pt}%
\noindent{\color{cream}\rule[-\figrulesep]{\columnwidth}{1.5pt}} }

\newcommand{\botfigrule}{\vspace*{-2pt}%
\noindent{\color{cream}\rule[\figrulesep]{\columnwidth}{1.5pt}} }

\newcommand{\dblfigrule}{\vspace*{-1pt}%
\noindent{\color{cream}\rule[-\figrulesep]{\textwidth}{1.5pt}} }

\makeatother
\twocolumn[
  \begin{@twocolumnfalse}
{
}\par
\vspace{1em}
\sffamily
\begin{tabular}{m{4.5cm} p{13.5cm} }

\includegraphics{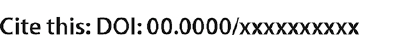} & \noindent\LARGE{\textbf{Impact of polyelectrolyte adsorption on the rheology of concentrated Poly(N-Isopropylacrylamide) microgel suspensions$^\dag$}} \\
\vspace{0.3cm} & \vspace{0.3cm} \\

 & \noindent\large{Rajam Elancheliyan,$^{\ast}$\textit{$^{a}$} Edouard Chauveau,\textit{$^{a}$} and Domenico Truzzolillo$^{\ast}$\textit{$^{a}$}} \\

\includegraphics{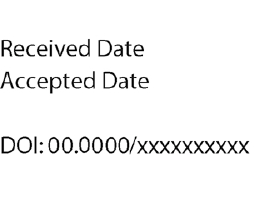} & \noindent\normalsize{We explore the impact of three water-soluble polyelectrolytes (PEs) on the flow of concentrated suspensions of poly(N-isopropylacrylamide) (PNIPAm) microgels with thermoresponsive anionic charge density. By progressively adding the PEs to a jammed suspension of swollen microgels, we show that the rheology of the mixtures is remarkably influenced by the sign of the PE charge, PE concentration and hydrophobicity only when the temperature is raised above the microgel volume phase transition temperature $T_c$, namely when microgels collapse, they are partially hydrophobic and form a volume-spanning colloidal gel.  
We find that the original gel is strengthened close to the isoelectric point, attained when microgels are mixed with cationic PEs, while PE hydrophobicity rules the gel strengthening at very high PE concentrations. Surprisingly, we find that polyelectrolyte adsorption or partial embedding of PE chains inside the microgel periphery occurs also when anionic polymers of polystyrene sulfonate with high degree of sulfonation are added. This gives rise to colloidal stabilization and to the melting of the original gel network above $T_c$. Contrastingly, the presence of polyelectrolytes in suspensions of swollen, jammed microgels results in a weak softening of the original repulsive glass, even when an apparent isoelectric condition is met. Our study puts forward the crucial role of electrostatics in thermosensitive microgels, unveiling an exciting new way to tailor the flow of these soft colloids and highlighting a largely unexplored path to engineer soft colloidal mixtures.} \\

\end{tabular}

 \end{@twocolumnfalse} \vspace{0.6cm}

  ]

\renewcommand*\rmdefault{bch}\normalfont\upshape
\rmfamily
\section*{}
\vspace{-1cm}


\footnotetext{\textit{$^{a}$~Laboratoire Charles Coulomb, UMR 5221, CNRS–Universit\'{e} de Montpellier, F-34095 Montpellier, France. Tel: +33 (0)467 143589; E-mail: domenico.truzzolillo@umontpellier.fr, rajam.elancheliyan@umontpellier.fr}}

\footnotetext{\dag~Electronic Supplementary Information (ESI) available: [\rv{Intensity autocorrelation functions, normalized time-averaged scattered intensities $I(q)/I_0$ and viscosimetry for dilute bare microgel suspensions. Rheology and rheological state determination at T=20 $^{\circ}$C and varying $\varphi$ of dense microgel suspensions. Rheological ageing of concentrated microgels at $\varphi=1.5$ and $\xi=0$}. Normalized mobilities and comparison table of electrophoretic mobilities for pure PE suspensions, bare microgels, and PE-microgel mixtures. See DOI: 10.1039/cXsm00000x/]}



\section{Introduction}
Colloid-polymer mixtures represent an ever-present paradigm for
manipulation of the microscopic dynamics and the flow properties of soft matter. They offer unique opportunities for addressing
challenges of central interest in the field of the glass transition and the gelation of colloids. Mixtures of neutral sub-micrometer particles and non-adsorbing linear polymers have been investigated in the last 20 years\cite{poon_physics_2002,pham_multiple_2002,vlassopoulos_tunable_2014}
showing that the rheology of colloidal suspensions can be drastically changed by depletion interactions of purely entropic nature\cite{binder_perspective_2014}. By contrast the effect of adsorbing polymers on the flow properties of colloids is much less explored. One emblematic case is the one of mixtures of charged colloids and polyelectrolytes, two types of macroions that are ubiquitus in nature.   
Polyelectrolytes (PEs) are charged polymers whose monomer segments bear an electrolyte group that dissociate in polar solvents. For this reason PEs are very sensitive to any electrostatic field and to the presence of other ionic species. They are widely employed to modify the properties of colloidal particle suspensions, membranes and solid surfaces\cite{szilagyi_polyelectrolyte_2014,bordi_polyelectrolyte-induced_2009}, and they are used in industry for many purposes ranging from water remediation to mineral separation and to control the rheology of particle slurries and pastes\cite{bolto_organic_2007,tobori_rheological_2003,howe_controlled_2006,pochard_effect_2010}.
PEs are also often employed to create protective and/or functional
coatings\cite{phenrat_stabilization_2008,salgueirino-maceira_coated_2003}, oppositely charged multilayers\cite{schwarz_polyelectrolyte_2012,decher_fuzzy_1997} or brushes by grafting or by adsorption of block copolymers\cite{schmidt_polyelectrolyte_2004,ballauff_polyelectrolyte_2006}. These coatings are employed, for example, to regulate surface properties, including wetting, lubrication and adhesion\cite{dalsin_protein_2005,elzbieciak-wodka_transfer_2011}.
The properties of colloid surfaces can be therefore remarkably changed by PEs addition and the understanding of the relationship between PE adsorption, particle interactions, and the
stability of the resulting mixtures is crucial for the future development of both polyelectrolyte additives and novel, soft and tunable materials.\\
The latter have attracted the attention of a large part of the scientific community working in materials science, and soft colloids have proved to be excellent constituents for their conception and designing.
Soft colloids like star polymers, microgels, micelles and vesicles, whose structure can be tailored at the molecular level\cite{vlassopoulos_tunable_2014}, are indeed model building blocks for materials with adjustable rheology and microscopic dynamics. Among them microgels made of poly(N-isopropylacrylamide)
(PNIPAm) are particularly interesting and intensely investigated
because they undergo a volume phase transition (VPT) at
ambient temperature: below $T_c\approx 32 ^{\circ}C$ these microscopic crosslinked networks are fully hydrated and swollen, while above $T_c$ they collapse due to their increased hydrophobicity. PNIPAm microgels can be synthesized using standard emulsion polymerization in aqueous media\cite{pelton_preparation_1986}, and since their first synthesis \cite{pelton_particle_1989} they showed intriguing electrostatic properties due to the presence of the ionic initiator used to promote chain polymerization. A drastic increase of microgel electrophoretic mobility has been reported for $T>T_c$ as a consequence of the large increase of their charge density driven by the particle collapse. This has raised many questions on the adsorbing power of PNIPAm microgels, especially when they are co-suspended with other charged species.
In this respect, some of the authors\cite{sennato_double-faced_2021} have recently pointed out the "double-faced" electrostatic behavior of PNIPAm microgels in aqueous media.
On the one hand, when microgels are swollen and oppositely charged polyelectrolytes or nanoparticles (Np) are progressively added, cluster formation does not occur in a wide range of PE or Np concentrations until when eventually salting out or other non-electrostatic effects destabilize the suspensions. On the other hand when microgels are in their collapsed state, they strongly interact with oppositely charged PEs or NPs, a large and sharp
mobility inversion occurs and large clusters form close to the isoelectric point, i.e., where the mobility of PE- or NP-microgel complexes is zero. This two-fold nature of PNIPAm microgels paves the way towards a temperature-sensitive complexation with charged polymers that might impact many potential applications including the controlled formation of micro-capsules \cite{prevot_behavior_2006,kim_one-step_2015} and membranes\cite{malaisamy_high-flux_2005}, gene delivery\cite{kabanov_interpolyelectrolyte_1998} and water treatment protocols\cite{bolto_organic_2007}, and that might also change drastically the rheology of fluid-fluid interfaces\cite{vialetto_influence_2022}. 
However, while the effect of polyelectrolytes\cite{sennato_double-faced_2021,truzzolillo_overcharging_2018,greinert_influence_2004,kleinen_defined_2008,kleinen_polyelectrolyte_2011,kleinen_rearrangements_2011} and simple ions\cite{lopez-leon_macroscopically_2007,lopez-leon_cationic_2006} on PNIAPm-based microgels has been very well detailed in literature, and both the rheology and the microscopic dynamics of concentrated bare microgel suspensions has been thoroughly investigated \cite{romeo_temperature-controlled_2010,sessoms_multiple_2009,philippe_glass_2018}, the effect of soluble PE addition in dense microgel systems and the role played by electrostatic and non-electrostatic adsorption still remain unknown. Only very recently mixtures of concentrated PNIAPm microgels and non-ionic surfactants have been investigated, unveiling a very rich phase diagram\cite{fussell_reversible_2019} and leaving open the question whether electrostatics has an important impact on the dynamics of the mixtures, especially at high temperatures where these colloids become densely charged.
In this work we want to elucidate this aspect by studying the effect of three types of known polyelectrolytes on the rheology of concentrated microgel suspensions, both below and above their critical temperature. We added separately two cationic and one anionic PEs with comparable molecular weights to suspensions of anionic PNIPAm microgels and study their linear rheology. The PEs have been chosen to vary both the sign of their charge and their hydrophobicity. We show that, while jammed suspensions of swollen microgels are weakly affected by the presence of PEs even when an apparent charge neutralization occurs, the rheology of collapsed and hydrophobic microgels is dramatically affected by PE addition and that both PE charge and hydrophobicity are important for the rheology of the mixtures. Electrophoresis and transmittance measurements allowed us to relate a large enhancement of the gel elasticity to the presence of concomitant charge inversion and reentrant condensation of microgels occurring in diluted suspensions.  
The rest of the work is organized as follows. In Section \ref{matmeth} we present the materials employed, we detail microgel synthesis and the techniques used to investigate PE-microgel mixtures. In Section \ref{resetdisc} we first present the result of a preliminary characterization of the pure polymers (microgels and PEs) via electrophoresis, light scattering, transmittance and rheology experiments. We then discuss the rheology of the mixtures and the complementary electrophoretic and transmittance experiments that allowed to rationalize our results. Finally, in Section \ref{conclusion} we make some concluding remarks, we summarize the key results and we put forward the perspectives of our work.

\section{Materials and methods}\label{matmeth}

\subsection{Microgel synthesis}

Poly(N-isopropylacrylamide) microgels were synthesized via emulsion polymerization\cite{senff_temperature_1999}. 150 mL of ultra-pure water was introduced in a 250 mL three-necked flask, and degassing was carried out using vacuum/argon cycles. Vacuum was achieved by a vane pump and the argon/vacuum sequence was repeated 6 times. Finally, argon was bubbled for 15 min. After having completed degassing, 3 to 4 mL of the (degassed) water was withdrawn via a syringe to dissolve 29.8 mg of the initiator (potassium peroxodisulfate, KPS – purchased from Sigma Aldrich and used without further purification) that was added in a later stage. Once the bubbling was stopped and the mechanical agitation was set up (via Teflon rotating anchor), the two side necks were plugged. At this stage 27.08 mg of Sodium Dodecyl Sulfate (SDS – purchased from Sigma Aldrich and used without further purification) was added in the flask right before the solution was heated up to the desired temperature ($T_s=70\pm 1$ $^{\circ}C$).
Once the target temperature was attained, 1.25 g of N-isopropylacrylamide (NIPAm) (from Sigma Aldrich, used without further purification) and 91.96 mg of N,N-methylene-bis-acrylamide (BIS) (from Sigma Aldrich, used without further purification) were introduced into the three-necked flask.  
During the heating ramp, the initiator (KPS) was dissolved in 4 mL of deionized and degassed water, and it was injected by hand slowly once the temperature of the batch reached $T_s$. The mixture was left under stirring at $T_s$ for 6 hours. The polymerization terminated spontaneously. All samples have been purified via three consecutive centrifugation/supernatant removal cycles \cite{conley_relationship_2019, del_monte_two-step_2021} and 19.5 mg (2 mM) of Sodium Azide have been added to prevent bacterial growth. 

After the purification step, the microgel suspensions were centrifugated to get a final microgel volume fraction $\varphi = 1.57$. Where $\varphi$ is the generalized volume fraction measured via rolling-ball viscosimetry (section \ref{visco}). We further diluted an aliquot of this sample to study the rheological behavior of pure PNIPAm suspensions at varying $\varphi$ and to prepare successively PE-microgel mixtures at fixed microgel volume fraction.     

\subsection{Polyelectrolytes (PEs)}\label{PEs}
\begin{figure}[htbp]
\centering
  \includegraphics[width=9cm]{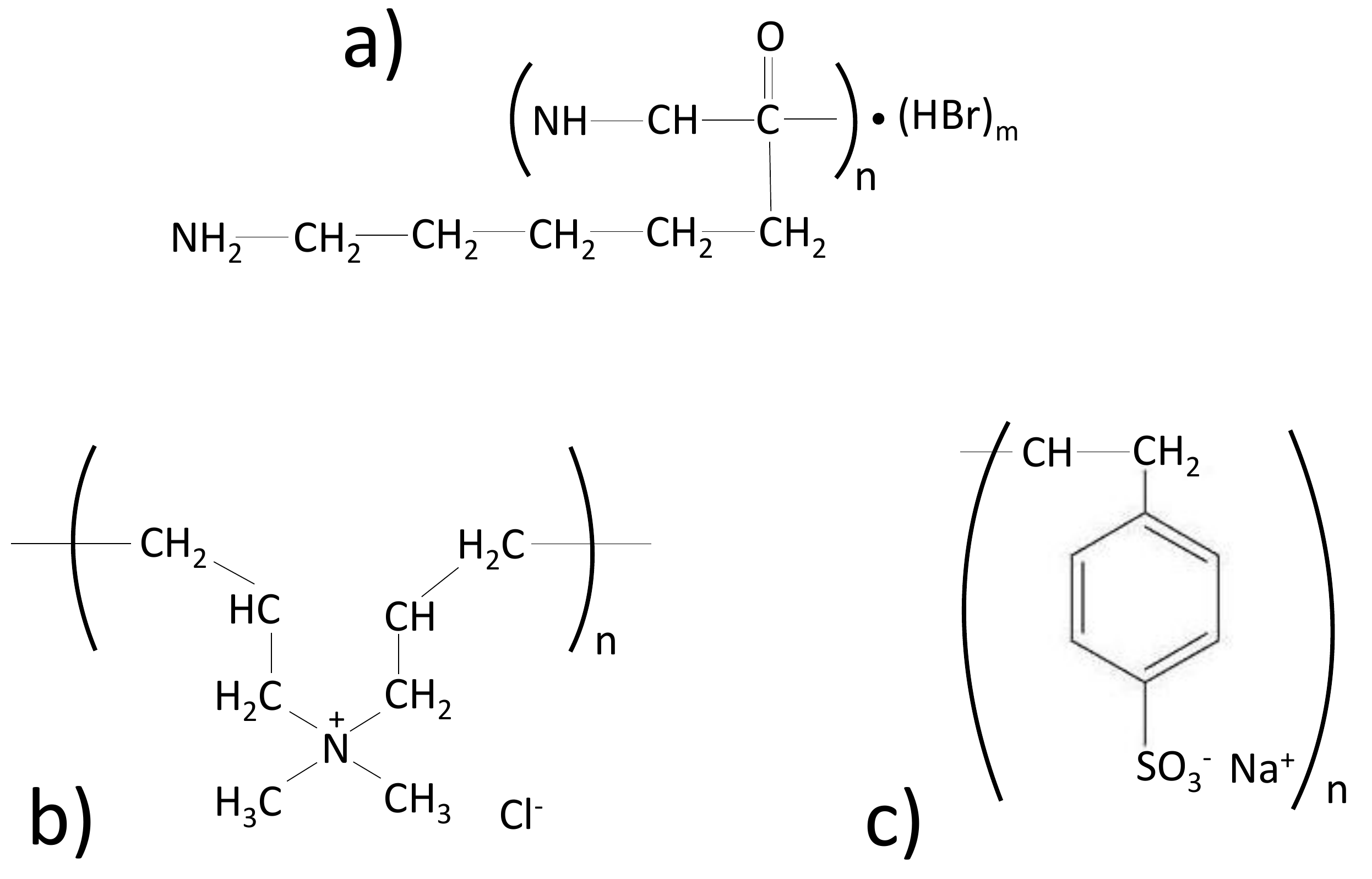}
  \caption{The structure formula of the polyions employed in this study: (a) $\alpha$-polylysine hydrobromide (PLL); (b) polydiallyldimethylammonium chloride (PDADMAC); (c) polystrirene sulfonate sodium salt (PSS)}
  \label{fig:chemical_PE}
\end{figure}

Cationic Poly-(l-lysine hydrobromide) (PLL) (Mw=50 kDa) and anionic Polystryrene sulfonate sodium salt (PSS) (Mw=43 kDa) were purchased from Polymer Source, Inc. (Canada). Cationic polydiallyldimethylammonium chloride (PDADMAC) (Mw<100 kDa) was purchased from Sigma Aldrich (Merck KGaA, USA). All PEs were used without further purification. They were dissolved in deionized salt-free water at varying concentrations and successively mixed with the microgel suspensions. The structure formula for the three repeating units of the PEs are shown in Figure \ref{fig:chemical_PE}.
The three polymers are characterized by different persistence lengths $l_p$ (stiffness) and hydrophobicity. In particular $l_p=$0.3 nm for PSS\cite{degiorgio_transient_1991}, $l_p=$1 nm for PLL\cite{shi_control_2013} and $l_p=$2.7 nm for PDADMAC\cite{mattison_complex_1998}. PLL is a weak polyelectrolyte\cite{burke_ph-responsive_2003} and it is the most hydrophobic polymer among those employed here, since each lysine bear a hydrophobic methylene side-chain that is responsible for chain association at high concentrations\cite{homchaudhuri_novel_2001,stagi_at_2022} and a tendency to penetrate into lipid membranes\cite{sennato_hybrid_2008}. Its degree of protonation depends on the pH, whose variation however stays very limited in M-PLL mixtures (see section \ref{mixtures}).
PSS is a strong polyelectrolyte whose hydrophobicity depends on its degree of sulfonation that in our case is high (90 \%), but not complete. Therefore we expect possible residual hydrophobic interactions between PSS chains and microgels, when the latter are in their collapsed state. Finally PDADMAC is a strong polyelectrolyte and it is the most hydrophilic polymer\cite{milyaeva_influence_2017} among those used in this work to investigate the rheology of PE-microgel mixtures, since it has neither large hydrophobic side chains as PLL nor neutral hydrophobic monomers on the backbone as PSS.  

\subsection{PE-Microgel suspensions}\label{mixtures}
\begin{figure}[htbp]
\centering
  \includegraphics[width=9cm]{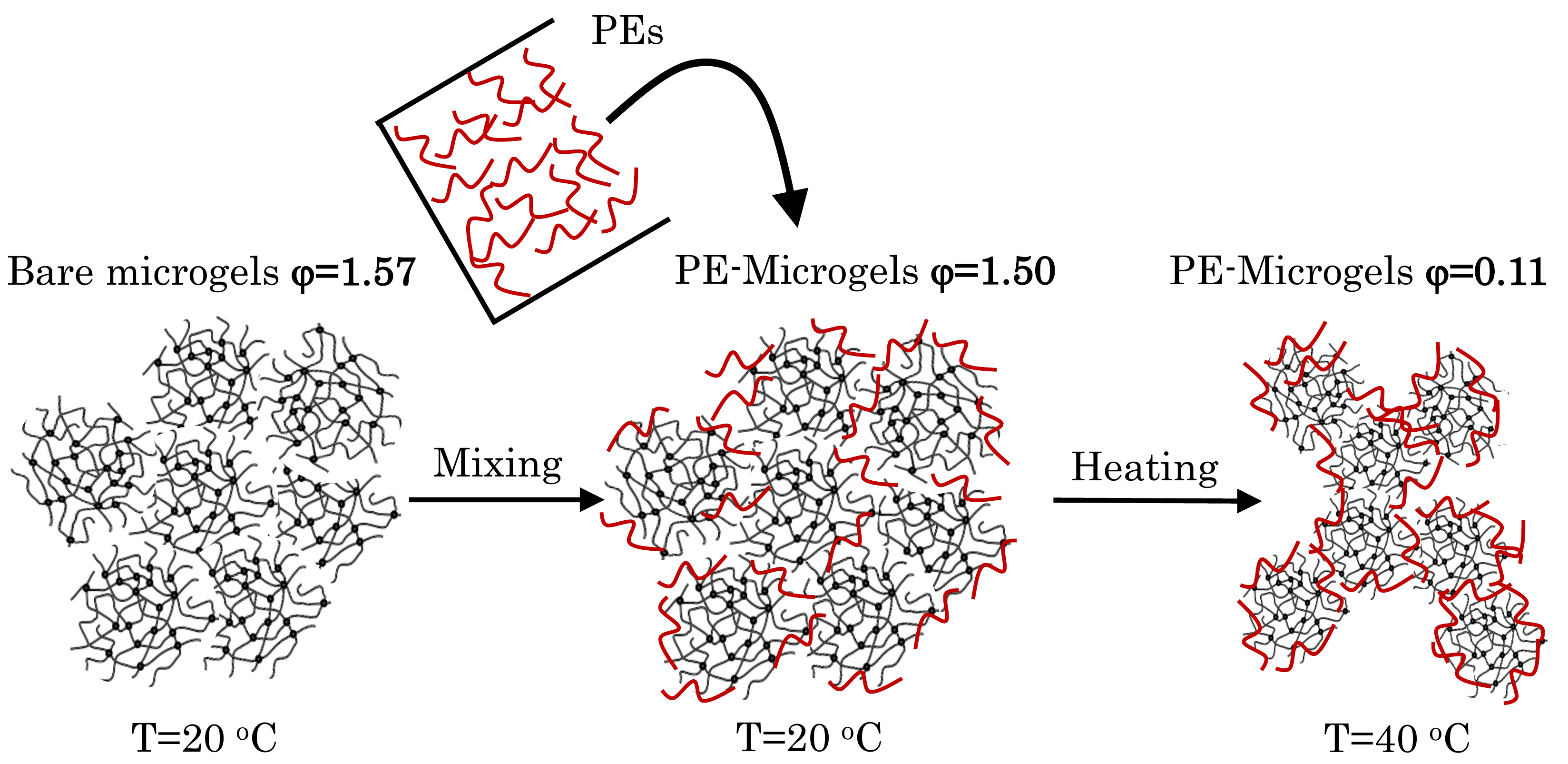}
  \caption{Schematic representation of the adopted protocol, including the mixing and the successive heating of the samples from T=20 $^{\circ}$C to T=40 $^{\circ}$C. Because of microgel deswelling the generalized volume fraction (section \ref{visco}) decreases from $\varphi(20^{\circ}C)=$1.5 to $\varphi(40^{\circ}C)=$0.11.}
  \label{protocol}
\end{figure}

Mixtures of polyelectrolytes and microgels, hereafter called PE-microgel mixtures and coded as M-PLL, M-PSS and M-PDADMAC, were prepared following the same protocol for both anionic (PSS) and cationic (PLL, PDADMAC) PEs: $23.33$ $\mu$L  of PE solution at the required concentration was added to 500 $\mu$L of microgel suspension at T=20 $^{\circ}$C with generalized volume fraction $\varphi =1.57$.
This protocol allows to get mixtures with fixed microgel volume fraction $\varphi =1.5$ and different concentrations of PEs. The mixtures were stirred for about 2 mins using vortex, and the resulting suspensions were then used for rheology measurements at T=20 $^{\circ}$C and T=40 $^{\circ}$C. Figure \ref{protocol} sketches the protocol, including the mixing and the successive heating of the samples in the rheometer geometry. 
We quantify the amount of PE contained in each suspension via the nominal polyelectrolyte to KPS monomolar ratio $\xi$, where KPS is considered as fully reacted during the synthesis. This is supported by the barely measurable weight of the residual mass ($m_r<$0.1 mg) present after drying the supernatant that has been extracted after each centrifugation. The monomolar ratio $\xi$ reads:
\begin{equation}\label{monoratio}
    \xi = \frac{C_{PE}\cdot M_w^{KPS}}{2\cdot C_{KPS}\cdot M_w^{PE}}
\end{equation}
where, $C_{PE}$, $C_{KPS}$, $M_w^{PE}$ and $M_w^{KPS}$ are respectively the concentrations (mg/ml) and the molecular weights of PE monomers and of the initiator.
One would expect to find an iso-electric point, where all the charges due to the anchored KPS are neutralized by the cationic PEs, to be close to $\xi=1$. However, we anticipate here that this condition will not be fulfilled, since not all the initiator molecules participate to build up the net charge of microgels\cite{elancheliyan_role_2022}.
The pH of the concentrated suspensions has been monitored showing in all cases only a weak dependence on the PE concentration: 5.5<pH<6.0 for M-PLL mixtures, 5.3<pH<6.0 for M-PDADMAC mixtures and 6.5<pH<6.0 for M-PSS mixtures. In this range of pH we do not expect any drastic change in PNIPAm microgel properties\cite{pei_effect_2004,al-manasir_effects_2009}. Most of mixtures have been further diluted 250 times and then used to measure the electrophoretic mobility of the diluted complexes and the fraction of the incident light transmitted through each sample at different temperatures.   
A small leftover volume of the the concentrated microgel suspension ($\varphi=$1.57) has been further mixed with PEs solutions to perform complementary mobility and light transmission experiments in the same range of $\xi$ explored via rheology.

\subsection{Viscosimetry}\label{visco}
Rolling-ball viscosimetry measurements were performed to obtain the generalized colloidal volume fraction of microgel suspensions \cite{senff_temperature_1999}.
The measurements were done at $T=$ 20 $^{\circ}$C  using an Anton Paar Lovis 2000 ME microviscosimeter in the range 4.68 $\cdot$ 10$^{-5}$ $<$ $c$(wt/wt) $<$  1.5 $\cdot$ 10$^{-3}$, where the viscosity $\eta$ increases linearly with the mass fraction of microgels $c$ (See Supplementary Material).
The particulate volume fraction is defined as $\varphi$ = $n_pv_p$, where $n_p$ is the particle number density and $v_p = 4\pi r^3/3$ is the volume of a single particle of radius $r$ at infinite dilution. Experimentally, only the concentration $c$ (wt/wt) of a (purified) suspension can be measured directly, by weighting a small volume of the sample before and after removing the solvent by evaporation.
Since the generalized volume fraction $\varphi$ is proportional to the mass concentration $c$, it can be replaced by $k\cdot c$, where $k$ is a factor for converting the mass concentration to the generalized volume fraction. We determined the constant $k$ matching the $c$ dependence of the zero shear viscosity $\eta$ of the purified suspensions with the values expected from the Einstein equation\cite{truzzolillo_overcharging_2018}:
\begin{equation}
   \frac{\eta}{\eta_s} =1 + \frac{5}{2}\varphi= 1 + \frac{5}{2}kc,
    \label{eq:Einstein}
\end{equation}
where $\eta_s$ is the viscosity of the solvent. By fitting $\eta/\eta_s$ to a straight line (see Supplementary Material), we thus obtained $k$, which allows extracting the colloidal volume fraction of the suspension.
We obtained $k=28.5$ $\pm$ $0.6$. This value embeds the effect of microgel permeability and that of the primary electroviscous effect as discussed recently \cite{elancheliyan_role_2022}.
Finally, we remind that for microgels with very similar synthesis \cite{philippe_glass_2018}  but lower crosslinker content, the onset of glassy dynamics at $T=20$ $^{\circ}$C occurs at $\varphi \approx 0.8$, that in our case represents an approximate lower bound for the liquid-to-solid transition. This threshold is compatible with our rheology data, as shown in sections \ref{bare}. 

\subsection{Light scattering}
Dynamic and static light scattering experiments have been performed to characterize the microgels at the single particle level.  
For this purpose an Amtec goniometer and a laser source ($\lambda$ = 532 nm) were used to collect the light at scattering angles in
the range 16$^\circ$ $\leq$ $\theta$ $\leq$ 150$^\circ$, corresponding to scattering wave vectors in
the range 4.4 $\mu m^{-1}$ $\leq$ $q$ $\leq$ 30.3 $\mu m^{-1}$. All scattering experiments have been performed in dilute samples: an aliquot of the purified mother batch was diluted in deionized water, to get a final generalized microgel volume fraction $\varphi=$0.006.

The hydrodynamic radius, $R_H$, and the polydispersity of the microgels were measured by means of dynamic light scattering. The scattered light intensity was collected at a fixed scattering angle ($\theta = 70^{\circ}$) correspondent to a scattering vector $q_{DLS}=18$ $\mu m^{-1}$ and analyzed using a digital autocorrelator. The time decay of the autocorrelation function $F_s(\vec{q},t)^2$ was then fitted by a second-order cumulant expansion (see Supplementary Material) to extract the diffusion coefficient $D$ as shown below \cite{frisken_revisiting_2001}:
\begin{equation}
   F_s(\vec{q}_{DLS},t)^2 \propto exp\left(-2q^2_{DLS}Dt\right) \left[1+\frac{\mu _2t^2}{2!}+ o(t^3)\right]^2
    \label{eq:cumulant}
\end{equation}
where $\mu_2$ is related to the second moment of the distribution of the diffusion coefficients of the suspended particles. The average diffusion coefficient is then used to obtain the hydrodynamic radius using the Stokes-Einstein relation: $D = K_BT/6 \pi \eta_s R_H $, where $K_B$ is the Boltzmann constant, $T$ is the bath temperature and $\eta_s$ is the zero shear viscosity of the solvent. The corresponding size dispersion and polydispersity index are respectively $\sigma_{R_H} = \sqrt{\mu_2}R_H/(Dq^{2}_{DLS})$ and $\gamma=\mu_2/D^2q_{DLS}^4$. The polydispersity index here never exceeded 0.20. For commercial PEs solutions, characterized by larger polydispersities indexes (PDI>0.2) autocorrelation functions have been analyzed by means of the CONTIN algorithm \cite{provencher_constrained_1982}  trough which we extracted the number-weighted size distributions.

The gyration radius, $R_g$, was measured by collecting the intensity of the light $I(q)$ scattered by the microgel samples at different scattering angles. The scattered light was subsequently fitted (see Supplementary Material) to the Guinier equation\cite{guinier_small_1955} to extract $R_g$:

\begin{equation}
   I(q) = I(0) \exp\left[-\frac{(qR_g)^2}{3}\right],
    \label{eq:guiner}
\end{equation}

where $I(0)$ is a constant depending on the number of particles in the scattering volume and on the scattering factor of a single particle. The Guinier regime for all samples was attained in the range $0.5 \le qR_g \le 2.5$, coherently with previously reported microgel syntheses \cite{gasser_form_2014, clara-rahola_structural_2012}. The uncertainty on $R_g$ is given by the fit error, the latter being less than $1.5\%$ of the best-fit value.

\subsection{Electrophoresis and Transmittance measurements}
The electrophoretic mobility and the transmittance of the suspensions were simultaneously measured using a Litesizer 500 (Anton Paar). The apparatus uses the new cmPASLS method, which is a recently developed PALS technology\cite{bellmann2019dynamic}. The absolute transmittance $T_A$, is computed as the ratio between the intensity of the light transmitted through the sample ($I$) and that of the incident beam ($I_0$): 
\begin{equation}
  T_A=I/I_{0},
    \label{eq:transmittance-abs}
\end{equation}
To filter out any effect due to solvent and the cell we computed the relative transmittance  
\begin{equation}
  T_R=T_A/T_A^{H_2O},
    \label{eq:transmittance}
\end{equation}
that is the ratio between the absolute transmittance of the suspension and that of the pure solvent (water here). This has been done for each set temperature.      

The electrophoretic mobility and transmittance were measured between 20 $^{\circ}C$ to 50 $^{\circ}C$ after a proper thermalization to monitor the effect of polyelectrolytes addition onto the mobility of swollen and collapsed microgels. The volume fraction of all the samples was fixed at $\varphi = 0.006$.

\subsection{Rheology}
Rheological tests were performed on freshly prepared PE-microgel mixtures using a stress-controlled MCR501 rheometer (Anton Paar, Germany). Standard stainless-steel sandblasted cone-plate geometry (25 mm diameter, 0.998$^{\circ}$ cone angle) has been used for all the tests. Temperature control has been ensured by means of a Peltier element (PTD- 200). The measuring temperatures were fixed at T=20 $^{\circ}$C and T=40 $^{\circ}$C. To ensure thermal equilibrium the sample has been kept at the desired temperature for 10 mins prior to each measurement. The outer rim of the samples has been covered with a low viscosity silicon oil (0.1 Pa s) to minimize evaporation. Dynamic strain sweep (DSS) tests were carried out before each dynamic frequency sweep (DFS) test to evaluate the extent of the linear regime, namely where the first-harmonic viscoelastic moduli  $G'(\gamma_0)$ and $G"(\gamma_0)$ do not appreciably change upon varying the strain amplitude. The lack of important ageing and the absence of evaporation were further tested for about the same duration of one experiment ($\sim$ 2500 s) at 20 $^{\circ}$C and 40 $^{\circ}$C via a time sweep experiment on one PE-free microgel suspension (Supplementary Material). This excluded that the variation of the moduli observed for the mixtures results from different ages of the pure microgel system. 
All frequency sweeps were done at strain amplitudes within the linear viscoelastic regime were the two moduli do not appreciably change for increasing $\gamma_0$.
DSS and DFS tests at 20 $^{\circ}$C started on freshly loaded samples after 10 mins of thermalization. The samples have been successively heated up to 40 $^{\circ}$C and, after other 10 mins of thermalization, a DFS test has been carried out.  Our rheological measurements therefore probe only the rapid formation, strengthening or melting of the original glassy or gel phases and they do not take into consideration possible coarsening processes that might occur over time scales much longer (several hours) than our experiment duration ($\sim$ 2500 s).  
 For liquid-like samples responding with torques well below the lower limits imposed by rheometer under oscillatory shear we performed steady rate experiments in the range 42 $s^{-1}$ $\leq\dot{\gamma}\leq$ 1000 $s^{-1}$ to measure their flow curves $\sigma(\dot{\gamma})$, to probe their Newtonian behavior and to extract possibly their zero-shear viscosity.   

\section{Results and discussion}\label{resetdisc}

\subsection{Bare microgels and PEs}\label{bare}
\begin{figure}[htbp]
\centering
  \includegraphics[width=9cm]{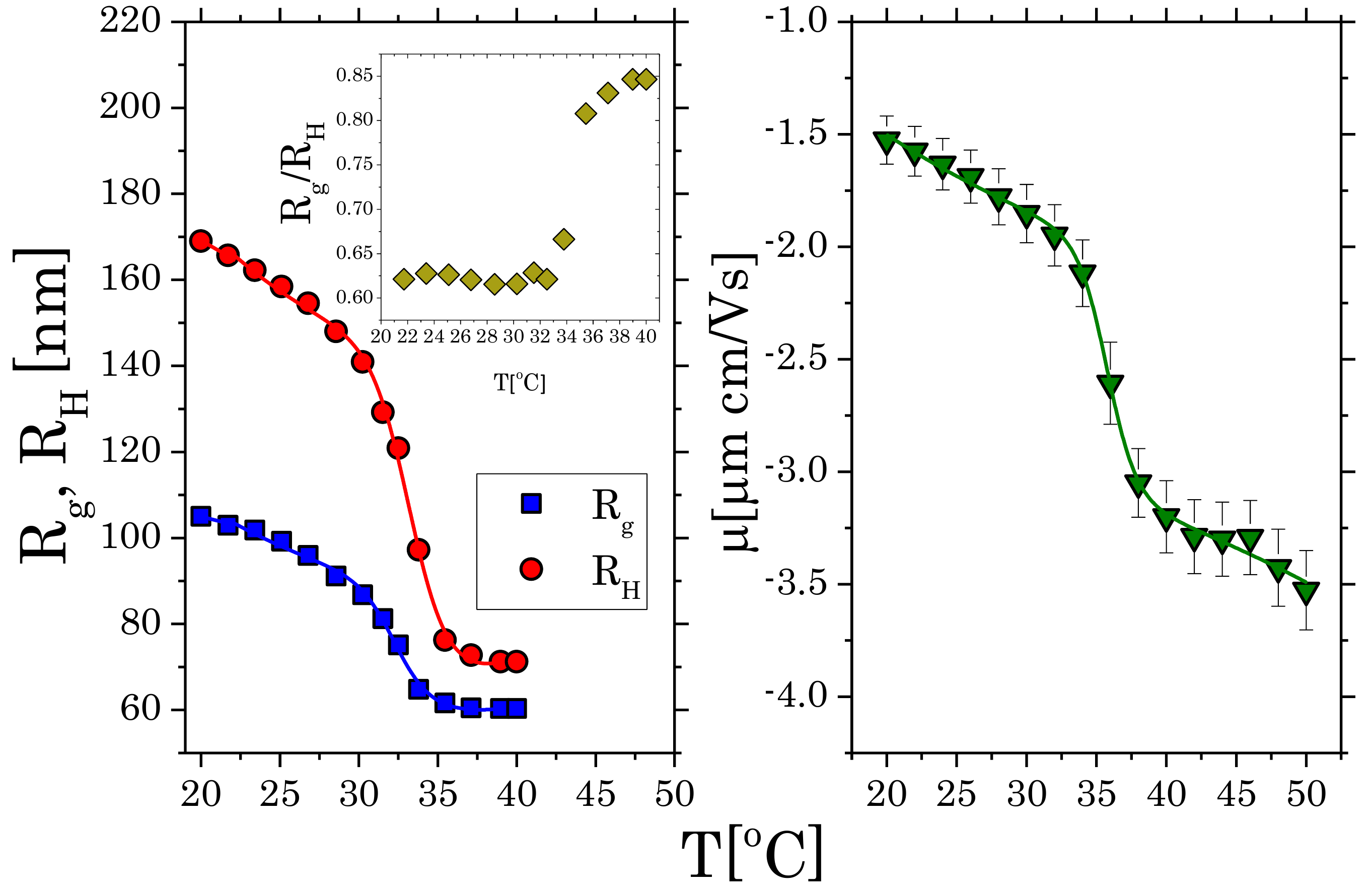}
  \caption{Gyration and hydrodynamic radius (left) and mobility (right) of bare PniPam microgels as a function of temperature. Here $\varphi=0.006$. The inset (left panel) shows the ratio $R_g/R_H (T)$. The relative error on $R_H$ and $R_g$ obtained from the fit of the intensity correlation function and the q-dependent scattered intensity (Equations \ref{eq:cumulant} and \ref{eq:guiner}) (See Supplementary Material) never exceeded the 1.5\% of the best fit values. Error bars for $R_H$, $R_g$ and $R_H/R_g$ are smaller than or equal to the symbol size. The error on the mobilities obtained from the full width at the half maximum of the mobility distribution never exceeded the 7\% of each mean value.}
  \label{fig:baremicrogels}
\end{figure}

Prior to mixture preparation we characterized the bare microgels and the PEs to confirm the sign of their charge in water, to measure their size and to identify rheologically the PE-free sample ($\xi=$0). 
The bare microgels were characterized using DLS and SLS in the dilute regime to measure their hydrodynamic radius ($R_H$),  gyration radius ($R_g$) and mobility $\mu$ as a function of temperature. As shown in fig \ref{fig:baremicrogels} (left panel) the microgels undergo a volume phase transition (VPT) when the temperature is raised above the LCST temperature of pNIPAM\cite{pelton_particle_1989}. 
The critical temperature ($T_c$) at which the microgels undergo the volume transition are estimated by fitting both $R_g$ and $R_H$ to an auxiliary function \cite{del_monte_two-step_2021}:
\begin{equation}
   R_{H,g}(T) = [R_0 - \Delta R_{H,g} \tanh(s(T-T_c^{H,g}))]+ A(T-T_c^{H,g})
    \label{eq:critical fit}
\end{equation}
where, $R_0$ is the radius of the microgel at the VPT, $\Delta R_{H,g}$ is the amplitude of the VPT and the parameter $s$ quantifies its sharpness. We obtained $T_c^H$ = 32.9 $\pm$ 0.2 $^{\circ}C$ and $T_c^g$ = 31.8 $\pm$ 0.3 $^{\circ}C$ for for $R_H$ and $R_g$ respectively. The lower $T_c$ for $R_g$ compared to $R_H$ can be attributed to uneven distribution of the charges between the core and periphery\cite{elancheliyan_role_2022, kleinen_defined_2008, truzzolillo_overcharging_2018}. 
Despite of that, given the dispersion of the data, we do not detect a clear onset of \rv{a} minimum of $R_g/R_H$ that is strictly related to the two-step deswelling of PNIAPm microgels\cite{del_monte_two-step_2021,elancheliyan_role_2022}, with the core collapsing at temperatures always lower than those marking the transition of the peripheral corona. In this respect we have already reported\cite{elancheliyan_role_2022} for a  synthesis of microgels with the same crosslinker-to-monomer molar ratio (5.3 \%) as the present one, the existence of a barely detectable minimum in $R_g/R_H$ crossing the VPT. \rv{Such a feature has to be ascribed} to the higher crosslinker density of the microgels employed here with respect to other syntheses where the minimum was more evident\cite{del_monte_two-step_2021}. 
\rv{As a matter of fact, increasing crosslinker density reduces the extent of the minimum of $R_g/R_H$ since it homogenizes the local deswelling within the microgel volume: the core and the corona deswell to the same extent for high crosslinker-to-monomer molar ratios and increasing temperatures, suppressing the decoupling between the transitions in $R_g$ and $R_H$ of the microgels. This has been carefully investigated and established via simulations\cite{del_monte_two-step_2021}.}
The ratio $R_g/R_H$ stays within the range 0.61-0.63 for $T\leq T_c^H$ and increases sharply at $T\simeq T_c^H$ consistently with the microgel shrinking above the VPT\cite{del_monte_two-step_2021}.     


\begin{figure}[htbp]
\centering
  \includegraphics[width=9cm]{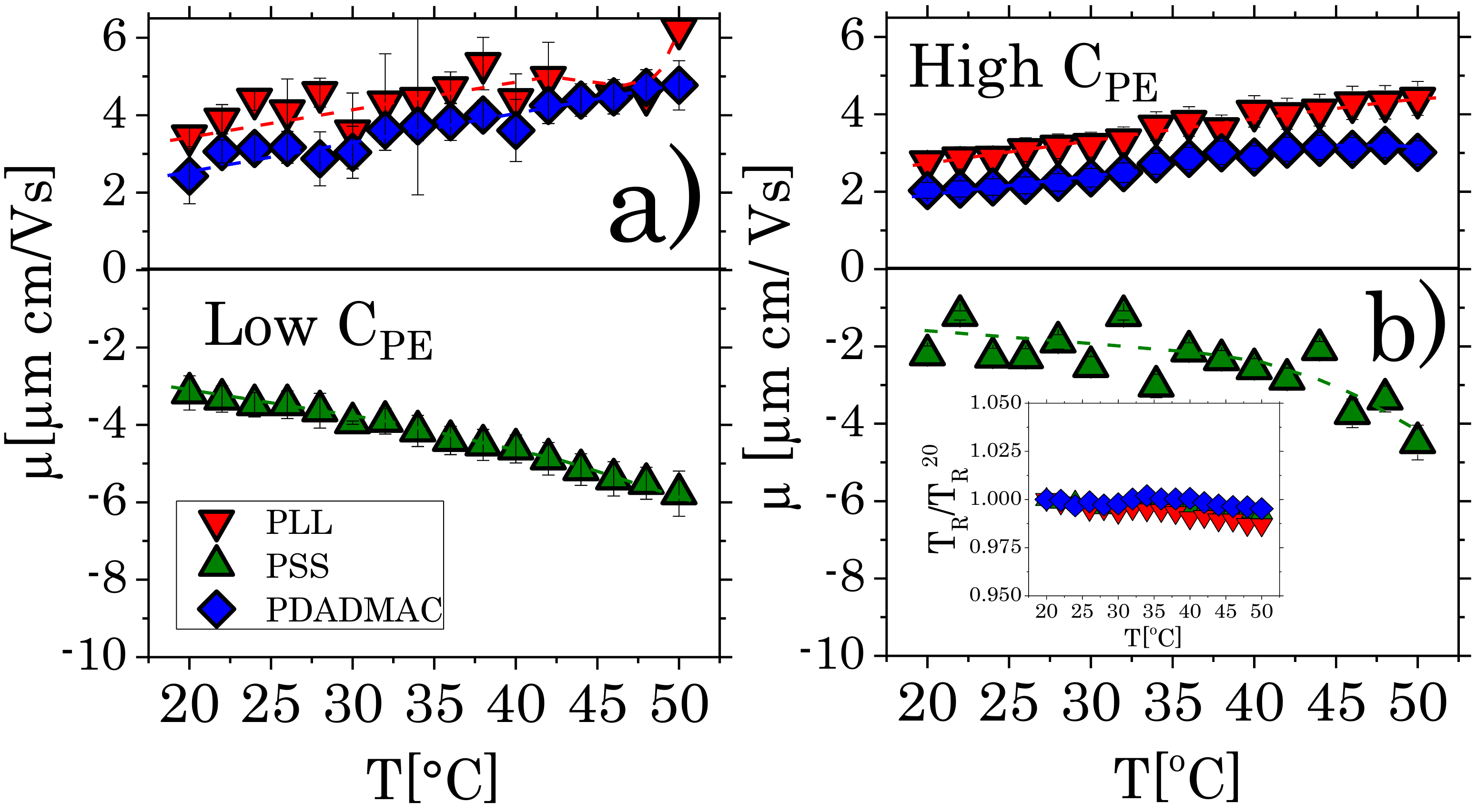}
  \caption{Panel a): Electrophoretic Mobility of the three PEs as a function of temperature at $C_{PE}$=1.25 mg/ml (for PLL), $C_{PE}$=1.75 mg/ml (for PSS) and $C_{PE}$=1.38 mg/ml (for PDADMAC). These concentrations are the highest ones (highest $\xi$) characterizing PE-microgel mixtures discussed in section \ref{PEM}. Electrophoretic Mobility at $C_{PE}$=28.0 mg/ml (for PLL), $C_{PE}$=5.1 mg/ml (for PSS) and $C_{PE}$=35.0 mg/ml (for PDADMAC). Error bars are the full widths at half maximum of mobility distributions. The inset (Panel b)) shows the normalized relative transmittance measured for the three concentrated PE samples as a function of temperature.}
  \label{fig:PEmob}
\end{figure}
The electrophoretic mobility of the same microgels are shown in Figure \ref{fig:baremicrogels} (right panel). The negative mobility is due to the anionic initiator used for the microgels synthesis. Microgel mobility is remarkably affected by the VPT\cite{pelton_particle_1989}: it drastically decreases (algebraically) above about $T_c^H$ due to the increase of the microgel charge density. We extracted the electrokinetic transition (EKT) temperature ($T_{c\mu}$) using the same auxiliary function as in equation \ref{eq:critical fit}, namely 
\begin{equation}
   \mu(T) = [\mu_0 - \Delta \mu \tanh(s_{\mu}(T-T_{c\mu}))]+ A_{\mu}(T-T_{c\mu})
    \label{mu:critical fit}
\end{equation}
where, $\mu_0$ is the mobility of the microgel at the EKT, $\Delta\mu$ is the amplitude of the EKT and the parameter $s_{\mu}$ quantifies its sharpness.
We obtain $T_{c\mu}$=35.8 $\pm$ 0.1 $^{\circ}C$. The discrepancy between the critical temperatures marking the VPT and EKT is $\Delta$=$T_{c\mu}$-$T_c^H\simeq$ 2.9 $^{\circ}C$. Such a significant difference between the transition temperature has been reported by Pelton \textit{et al.}\cite{pelton_particle_1989}, Daly et al.\cite{daly_temperaturedependent_2000} and more recently by Truzzolillo \textit{et al.}\cite{truzzolillo_overcharging_2018}, and it is attributed to a further charge restructuring (densification) well above $T_c^H$\cite{daly_temperaturedependent_2000}.    

\rv{To determine preliminarily the viscoelastic properties of the PE-free microgel suspensions, the rheology of bare microgels dispersions was also investigated as a function of $\varphi$ at T=20 $^\circ$C. This allowed us to determine precisely the rheological state of the suspension in which PEs are progressively added: At the microgel concentration ($\varphi=1.5$) that will characterize all the PE-microgel mixtures, microgels are in a jammed glassy state at T=20 $^{\circ}$C. (see Supplementary Material for more details).}
Prior to mixture preparation we have further investigated the temperature dependence of PE mobility and, since PLL and PSS bear also hydrophobic segments, we also inspected the PE stability via light transmission experiments at high concentration. Figure \ref{fig:PEmob}-a shows the electrophoretic mobility of the three PEs at the highest concentrations of the range explored in rheology experiments. As expected the 2 cationic PEs (PLL and PDADMAC) show positive mobility while PSS chains are characterized by negative mobilities at all temperatures. For the three PEs we observe a smooth increase of the mobility modulus with increasing temperature. Such an increase is mainly due to the decrease of the solvent viscosity (see Supplementary Material). Figure \ref{fig:PEmob}-b shows the mobility and the normalized transmittance of solutions that are more concentrated in PE chains. We used these samples to test the stability of all the PEs in water, reduce the uncertainties on the mobility measurements and test the effect of PE concentration on the mobility itself. By increasing PE concentration we observe a clear decrease in mobility for all the polymers, that is consistent with an increased fraction of condensed counterions\cite{truzzolillo_counterion_2009} and possibly an augmented friction due to more frequent collisions among different chains, while the observed temperature dependence can be still filtered out by taking into account the viscosity variation of the solvent (see Supplementary Material). Most importantly, the normalized transmittance of the same samples, namely the transmittance measured at temperature $T$ divided by its value at 20$^\circ$C (inset of Figure \ref{fig:PEmob}), does not vary remarkably when temperature increases and the non-normalized relative transmittance was $T_R$(20$^{\circ}$C)= 1.00$\pm$0.01 for the three PE solutions. 
The polyions employed here stay therefore well dissolved in water and any possible conformational change, \rv{local chain association}, or variation of the fraction of free counterions \cite{de_influence_2015,karpov_hydration_2016,hu_temperature-dependent_2020}  do not cause any massive condensation or chain swelling. This is an important starting point since transmittance will allow us to discern whether colloidal condensation occurs or not in the mixtures. To establish whether polyions can penetrate microgels we have finally measured or estimated the size of the PEs and the average mesh size of the microgels, the latter being equal to the average distance between two crosslinker molecules thought as uniformly distributed within the microgel volume. The hydrodynamic size distribution of PSS and PDADMAC chains has been obtained via the CONTIN analysis of the intensity autocorrelation functions \cite{provencher_constrained_1982}. DLS experiments have been performed using dilute PE suspensions, namely for $C_{PE}\ll C_{PE}^*=\frac{M_w}{4/3\pi Na R_H}$, where $C_{PE}^*$ is the overlap concentration and $N_A$ is the Avogadro number. We obtained number-weighted size distributions with a main peak at hydrodynamic diameters $D_H^{peak}=4.5$ nm for PSS and $D_H^{peak}=184$ nm for PDADMAC. 
We attribute this large difference in size between these two PEs to both a different average molecular weight and a known large difference in backbone stiffness (see section \ref{PEs}).
PLL chains did not give enough scattering signal for their hydrodynamic size to be measured reliably in the dilute regime. We have however estimated their average end-to-end distance by considering a worm-like chain model for semiflexible chains\cite{wang_simulation_2015,jin_investigating_2014} and the known persistence length of PLL, $l_p=1.0$ nm \cite{shi_control_2013}. We obtained an end-to-end distance $R_{ee}=28$ nm. The overlap concentration for PLL chains has been then estimated by replacing the hydrodynamic radius with half of the end-to-end distance.
The average mesh size $d_m$ of PNIPAm microgels has been computed by knowing the amount BIS molecules and the number of microgels contained in the mother batch. The former is known from the synthesis, while the latter can be computed knowing the value of the generalized volume fraction obtained by viscosimetry at T=20 $^\circ$C, the hydrodynamic size of microgels at the same temperature and the total volume $V$ of the suspension. We obtained $d_m=R_H[4\pi/(3 N_c)]^{1/3}=$ 4.9 nm, where $N_c$ is the number of crosslinkers per microgel. Since $d_m$ is always comparable to or lower than the measured or estimated size of the PEs \rv{the penetration of PE chains} is limited to the outer shell of the microgels where a lower than average crosslinker density characterizes PNIPAm microgels synthesized via free radical polymerization\cite{ninarello_modeling_2019,stieger_small-angle_2004}. \rv{We further expect therefore that the PE penetration is maximum for the small PSS chains and nearly absent for the large PDADMAC polymers. 
Finally, PE diffusion within the microgel volume is supposed to be additionally reduced at $40$ $^\circ$C, where microgels collapse and their mesh size consequently decreases.}

\subsection{Rheology of PE-microgel mixtures}\label{PEM}
\begin{figure}[ht]
 \centering
 \includegraphics[width=9cm]{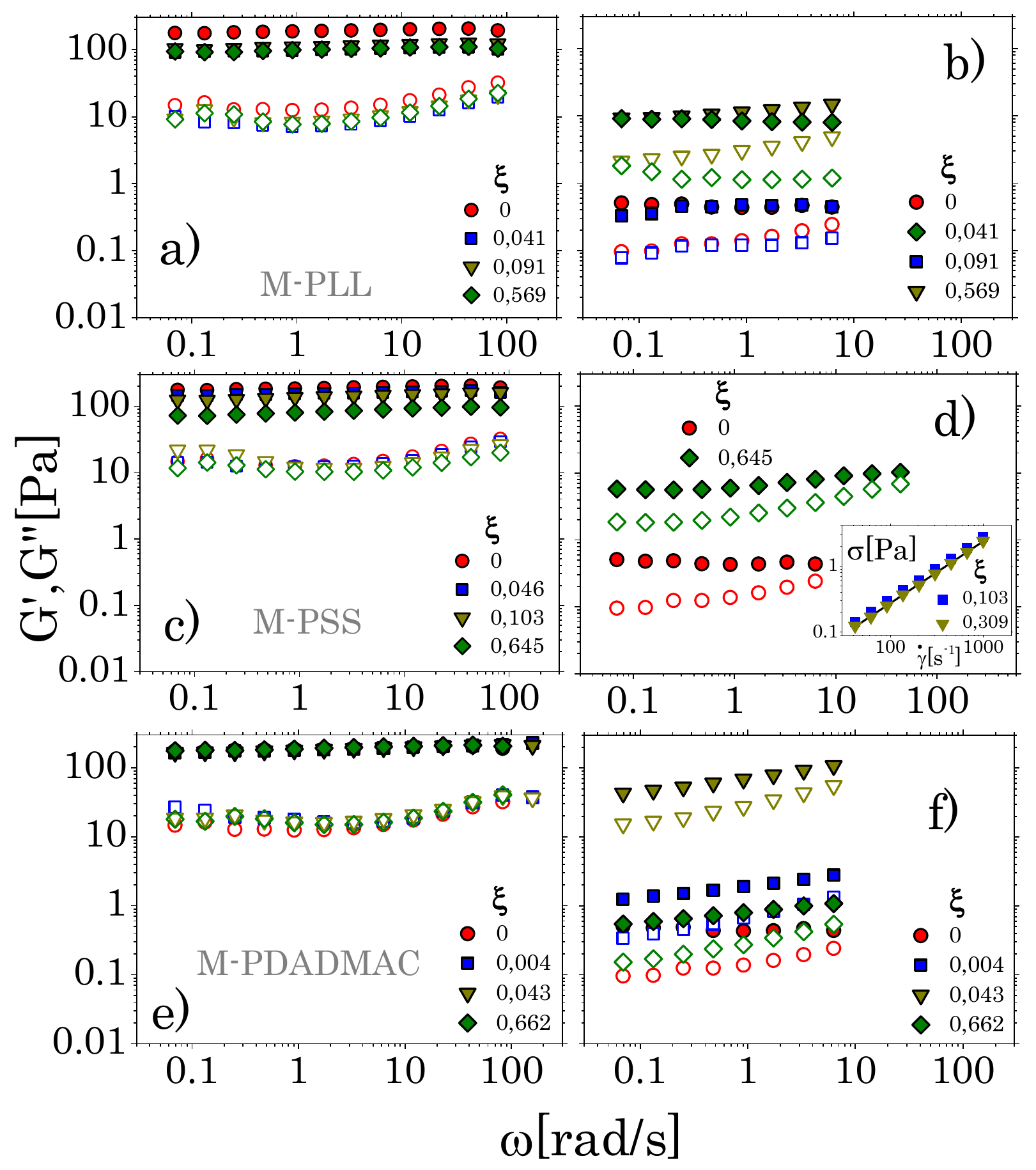}
 \caption{Storage modulus G' (solid symbols) and loss modulus G" (open symbols) as a function of the oscillatory frequency $\omega$ at T=20$^{\circ}$C (a,c,e) and T=40$^{\circ}$C (b,d,f) for M-PLL mixtures (a,b), M-PSS mixtures (c,d) and M-PDADMAC mixtures (e,f). The inset in panel (d) shows the flow curves $\sigma(\dot{\gamma})$ of liquid samples for which LVE moduli were not measurable. The straight line is a linear fit of the data at $\xi=$0.309.}
 \label{fig:DFS2}
\end{figure}

Fig \ref{fig:DFS2} shows selected dynamic frequency sweeps for the three sets of mixtures and different charge ratio $\xi$ at 20 $^{\circ}$C (a,c,e) and 40 $^{\circ}$C (b,d,f), namely below and above the microgel VPT. 
\rv{The limited range of frequency at 40 $^{\circ}$C is due to inertia problems, that are routinely encountered at high frequencies ($\gtrsim$ 10 rad/s) and standard shear geometries like the one given by the cone-and-plate fixture used in this work, producing a non-physical drop of the moduli for ultra soft solids ($Gp\lesssim$ 10 Pa).}
At 20 $^{\circ}$C the addition of PEs has only a weak effect on the rheology of the suspensions: we observe a weakining of the original jammed glass for M-PLL (a) and M-PSS (b) mixtures for increasing $\xi$, while the linear viscoelastic spectra of the mixtures stay basically unaltered when PDADMAC is added (c). We attribute this to 3 synergistic effects: i) microgel deswelling due to an increased ionic strength \cite{rasmusson_flocculation_2004} and osmotic pressure exerted possibly by unadsorbed chains \cite{saunders_thermal_nodate}; ii) depletion interactions that may act in presence of free chains, especially in the case of anionic PSS polymers \cite{snowden_temperature-controlled_1992}; iii) a reduction of the net repulsions between microgels due to a partial adsorption of the cationic polymers that lowers the electrostatic repulsion between microgel coronas bearing most of the microgel charge \cite{del_monte_two-step_2021}. As said, however, the effect PDADMAC chains on swollen microgel glasses does not emerge. In this respect it is worth recalling that PDADMAC chains showed a lower mobility at all temperatures (Figure \ref{fig:PEmob}-a,b) compared to PLL pointing to a lower charge density on the polymer backbone. We thus expect a lower adsorption energy and a weaker impact on the swollen microgels with low charge density. Finally, further effects might be produced by the PE polydispersity that we expect to be larger for PDADMAC as indicated by the supplier and by the different microgel-to-PE size ratio, \rv{since they both affect residual depletion effects}. \rv{In particular, since at T=20 $^{\circ}$C the hydrodynamic PDADMAC/microgel size ratio is $\simeq$ 1.08, while the (estimated) hydrodynamic PLL/microgel size ratio is $\simeq$ 0.082 and the adsorption is limited by the low charge density and high hydrophilicity of the microgels, a more prominent depletion (short-ranged) attraction might characterize their mutual interaction in presence of PLL rather than of PDADMAC chains, the latter being almost equally sized with respect to the PNIPAm microgels. In addition to that, depletion effects in M-PDADMAC mixtures are further reduced due to the presumable higher polydispersity of the depletants (the PEs here), whose impact on entropy driven interactions have been discussed in detail\cite{chuEffectsParticleSize1996}. }
\rv{This said}, despite these effects are difficult to quantify or decouple, they all point towards a softening of the colloidal glasses due to a reduction of the repulsive forces between the microgels.       
\begin{figure*}
 \centering
 \includegraphics[height=10.5cm]{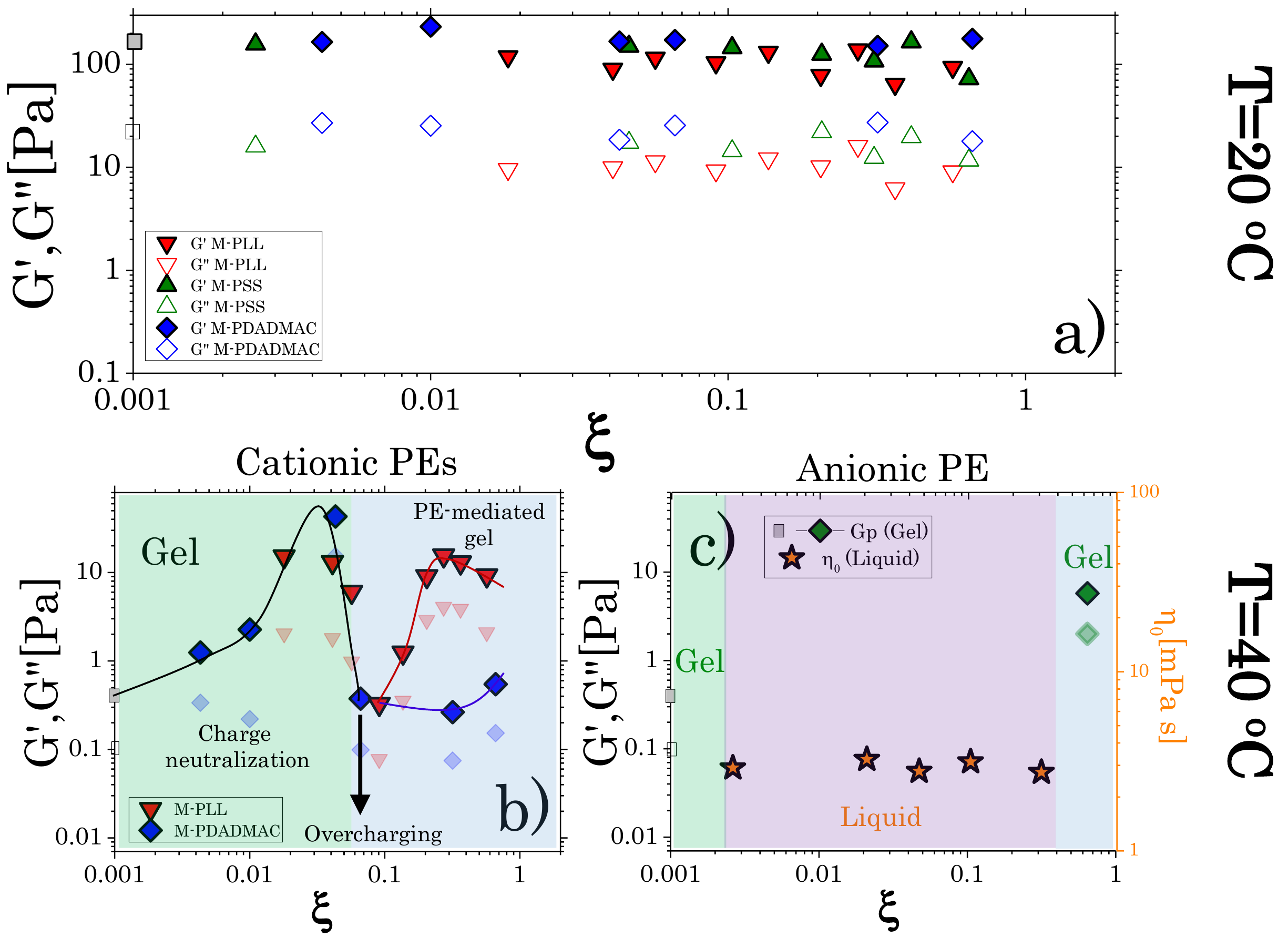}
 \caption{Storage (full points) and loss modulus (empty/shaded points) of PE-microgel mixtures for the three PEs as a function of monomolar ratio $\xi$ (eq. \ref{monoratio}) at $\omega=$0.07 rad/s, T=20$^{\circ}$C (a) and T=40$^{\circ}$C (b,c). For M-PSS mixtures the zero shear viscosity is also shown (c). The moduli of PE-free suspensions (grey squares) are also shown as reference. Solid lines in panel b) are a guide to the eye.}
\label{fig:dfs40}
\end{figure*}
At 40 $^{\circ}$C by contrast, the emerging scenario is very different and the rheology of microgel suspensions become very sensitive to both PE concentration and sign of the PE charge (Figure \ref{fig:DFS2}-b,d,f). First of all, we point out that at this temperature PE-free microgel suspensions ($\xi=0$) are still viscoelastic solids characterized by a nearly frequency-independent storage modulus $G_p=0.47\pm 0.05$ Pa. The generalized microgel volume fraction in this case is much lower than its value at 20 $^{\circ}$C, since microgels are in their collapsed state. By rescaling $\varphi$ using the the hydrodynamic radii measured at the two temperatures we obtain $\left.\varphi\right|_{40 {^\circ} C}=\left.\varphi\right|_{20 {^\circ} C}\frac{R_H(40 {^\circ} C)^3}{R_H(20 {^\circ} C)^3}=0.11$. At this volume fraction the suspensions cannot be considered any more as jammed glasses but they are rather in a gel phase, consistently with other rheological studies on concentrated PNIPAm suspensions \cite{romeo_temperature-controlled_2010}. Microgels aggregate and form percolating networks due to hydrophobic forces acting between the particles for $T>T_c^H$. \rv{It's worth remarking that this occurs despite of the temperature-induced increase of the charge density of the microgels, that stabilizes the suspensions at lower volume fractions, and it is caused by both the high number density of counterions, that screen microgel charges, and the large number density of microgels, that enhances the collision probability between two of these colloids, increasing the rate of cluster formation. The impact of the microgel charge and the counterion concentration on microgel gelation has been investigated only very recently\cite{minamiCriteriaColloidalGelation2020}, confirming that the suppression of electrostatic repulsions gives rise to a net increase of gel elasticity.} As PE concentration is progressively increased we observe three interesting phenomena: i) the presence of a large gel strenghtening followed by a sharp softening for both the sets of mixtures containing cationic PEs (M-PLL, M-PDADMAC); ii) a complete melting of the gel in wide range of PE concentrations as the anionic PSS chains are progressively added;
iii) a clear gel strengthening or re-gelification at the largest PE concentrations in M-PLL and M-PSS mixtures.  
For liquid-like M-PSS mixtures, since the linear viscoelastic moduli were not measurable, we performed continuous steady-rate tests and we extracted their flow curves (Inset Figure \ref{fig:DFS2}-d). They all showed a Newtonian behavior $\sigma(\dot{\gamma})=\eta_0\dot{\gamma}$, from which we obtained the zero shear viscosity $\eta_0$.       

Figure \ref{fig:dfs40} summarizes our results for all the samples investigated via rheology and serves to detail more clearly the effect of PEs. We report both the storage and the loss modulus of the solid-like suspensions at $\omega=0.07$ rad/s and the zero shear viscosity for the liquid-like samples. Panel (a) shows the moduli of all the mixtures as a function of $\xi$ at T=20 $^{\circ}$C: Increasing $\xi$ does not affect remarkably the rheology of the mixtures and produces only a weak decrease of both the moduli in M-PLL and M-PSS mixtures, while the effect of PDADMAC addition is not even detectable. Panels (b) and (c) by contrast show the relevant impact that PEs have on microgel networks. 
On the one hand both M-PLL and M-PDADMAc mixtures (b) are characterized by moduli that increase by almost two order of magnitudes with respect to the original gel ($\xi=0$) at $\xi\simeq 0.04-0.05$. The same moduli decrease again sharply at $\xi \simeq 0.1$. At higher PE concentration ($\xi > 0.1$) the two cationic PEs act differently on the network dynamics: the gel moduli increase again in M-PLL mixtures while this does not occur in M-PDADMAC suspensions. On the other hand, in stark contrast with the phenomenology encountered by adding oppositely charged PEs, the gel network is melted by PSS chains even at the lowest $\xi$, with the mixtures becoming Newtonian liquids with low viscosities ($\eta_0\simeq$3 mPa s). Also in this case, however, a large PE content destabilize the suspensions and a gel-phase is observed again.
The rheology of the binary PE-microgel mixtures therefore point to the existence of an emergent reentrant behavior of the gel at high temperature when cationic PEs are added, suggesting that PEs may first screen completely the entire microgel charge and then overcompesate for it, producing globally an overcharge of the PE-microgel complexes so formed. This is in line with the findings that some of the authors \cite{sennato_double-faced_2021} obtained in dilute suspensions of similar M-PLL mixtures and it will be confirmed by mobility and light transmittance experiments discussed later (section \ref{mob-et-trans}). The further gel strengthening occurring at high PLL content ($\xi>$0.1) and its absence for PDADMAC points to a key role played by PE hydrophobicity. PLL chains are more hydrophobic than PDADMAC, \rv{they tend} to aggregate at high concentrations as mentioned in section \ref{PEs} \rv{and they can thus participate actively to the enhanced elasticity of the gel}. In this respect we point out that for M-PLL mixtures the PE concentration reaches values above the overlap concentration $C_{PLL}^*$, spanning the range $0\leq C_{PLL}\leq 4.1 C_{PLL}^*$. Our \rv{results thus suggest} that, for high PE content, gels are at least partially PE-mediated: the gelation is driven also by the PE-PE interaction rather than only the PE-microgel one. We recall \rv{also} that both the cationic polymers progressively adsorb on the microgels at 40 $^{\circ}$C (see also section \ref{mob-et-trans}) and depletion effects \rv{should be considered of secondary importance}.\\    
The completely different behavior observed in M-PSS mixtures points quite surprisingly to a strong influence of polystyrene sulfonate on the state of the microgel suspensions, since PSS has been considered as a non adsorbing polymer for anionic PNIPAm microgels\cite{rasmusson_flocculation_2004}. We stress once again the PSS chains used in this work consist of 90 \% sulfonated polystyrene monomers, that is to say, polymer backbones having 1 hydrophobic styrene monomer over 10. This puts forward that PE adsorption might occur also in this case. In addition to this, since the mixtures have been prepared at T=20 $^{\circ}$C, i.e. at $\varphi=1.5$, PSS chains might partially penetrate microgels for entropic reasons, and after microgel collapse part of them can be still confined within the microgel volume. 
PSS adsorption or inclusion would explain the gel melting, since sulfonated chains are globally hydrophilic and their layer on microgels increase the negative charge of the M-PSS complexes hampering gelation. 
\rv{As for partially hydrophobic PLL chains, PSS causes a re-condensation of microgels at high concentrations ($\xi>$0.31).
Enhanced clustering in this case is only driven by hydrophobic interactions since there cannot be charge patch attraction\cite{velegol_analytical_2001} characterizing M-PLL and M-PDADMAC systems.
In more detail, we attribute this microgel aggregation to} i) concomitant energetic bridging due to the residual hydrophobic interactions between non-sulfonated styrene monomers and those between hydrophobic PNIPAm segments, ii) a simple salting out effect and iii) possibly depletion attractions due to non-adsorbed chains at high concentrations. Although our mobility measurements, that we discuss hereafter, confirm the (at least partial) inclusion of PSS chains within the microgel volume, a more systematic study \rv{involving other complementary techniques and simulations}, that goes beyond the scope of this work, is in order to decouple all these effects.
In what follows, we report on detailed electrophoretic and light transmission experiments in diluted mixtures, through which we have unravelled the role of the sign of the PE charge, and investigated the presence of an isoelectric point and colloidal overcharging causing a reentrant condensation of PE-microgel complexes in the same range of charge ratios probed by rheology experiments.   

\subsection{Mobility and light transmission experiments for diluted mixtures}\label{mob-et-trans}
\begin{figure*}[ht]
 \centering
 \includegraphics[height=10.40cm]{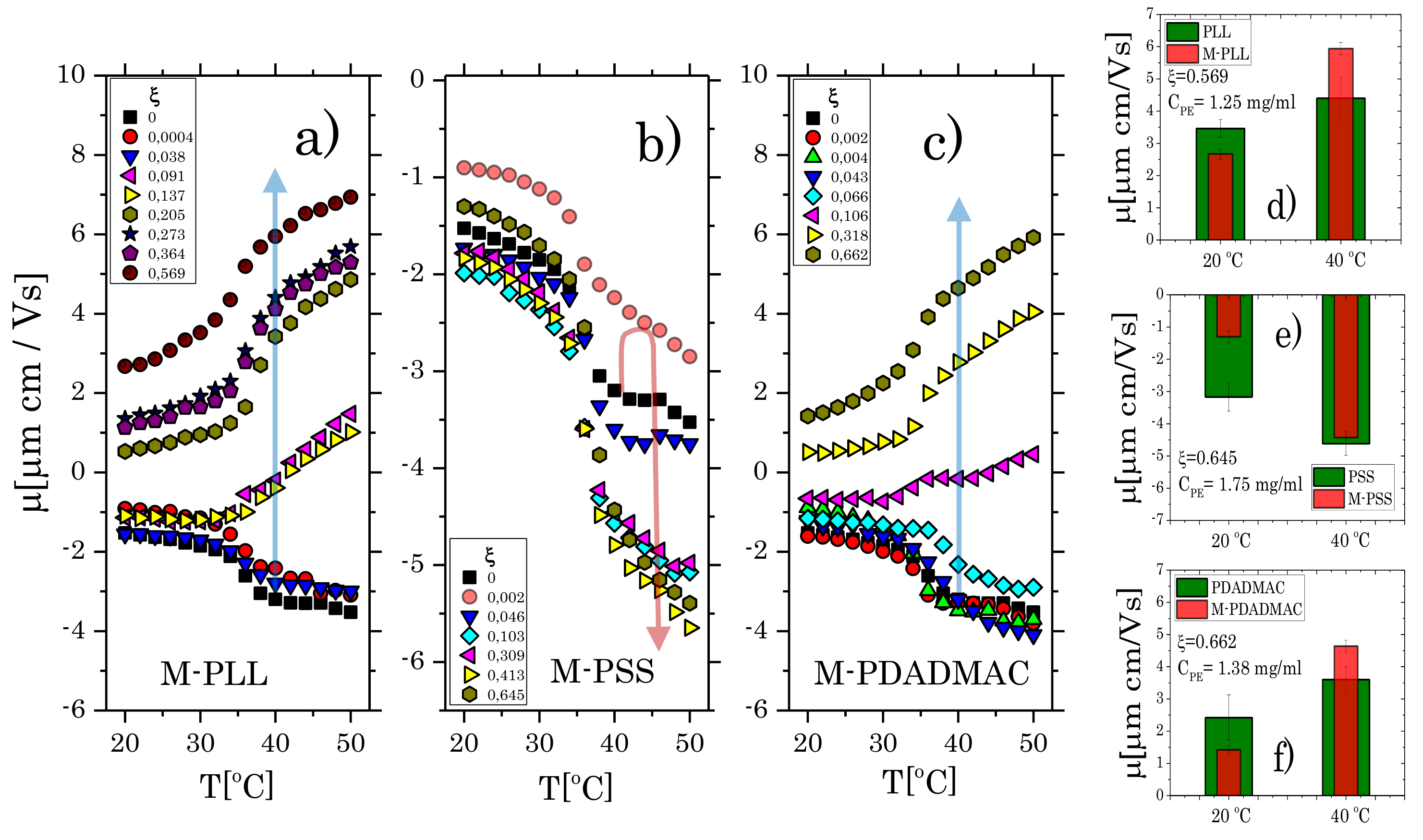}
 \caption{Electrophoretic mobility of a) M-PLL, b) and M-PSS and c) M-PDADMAC complexes at varying PE concentrations as a function of temperature. The arrows are a guide for the eye and point to the mobility variation at $T>T_c^H$ from $\xi=$0 to the maximum PE concentrations. Panels d), e) and f) show the mobility values for the larges PE concentrations in the mixtures compared to the values measured in pure polyelectrolyte solutions at the same concentrations as indicated in the panels.}
 \label{fig:tvsmob}
\end{figure*}
Most of the suspensions studied via rheology have been diluted in deionized water so to obtain PE-microgels mixtures 
with generalized microgel volume fraction $\varphi=0.006$ at T= 20 $^{\circ}$C. Figure \ref{fig:tvsmob} shows the elctrophoretic mobility for the three systems as a function of temperature. The behavior of $\mu(T)$ changes drastically depending on whether the cationic PLL and PDADMAC polymers (Panels a,c) or anionic PSS (Panel b) are progressively added, staying qualitatively unaltered when the effect of the temperature dependence of the viscosity and the permittivity of the solvent is filtered out (Supplementary Material). For both the cationic PEs we observe a pronounced overcharging starting at $\xi\simeq 0.1$ that increase radically at $T>T_{c\mu}$. Such overcharging has been already observed \cite{truzzolillo_overcharging_2018,sennato_double-faced_2021} and it is driven by the microgel EKT, with the largest increase of mobility occurring close to the native electrokinetic transition of PNIPAm microgels. This will be briefly discussed below.   
The mobility behavior for the M-PSS reveals by contrast an unexpected behavior: after a first algebraic increase, presumably due  to an increase of the ionic strength and a scarce PE adsorption, the mobility increases in modulus at all temperatures almost doubling its magnitude at T=50 $^{\circ}$C. This strongly indicates that PSS chains adsorb onto microgels or stay confined within their peripheral volume. We must conclude that part of the PSS chains adsorb, potentially penetrating inside the microgel periphery, and that the hydrophobic interactions between PSS chains and microgels, due to the not fully sulfonation of polystyrene chains, might favor PE coating despite of the anoinic nature of the PE. 

Such mobility change does explain why microgels are stabilized by PSS chains in concentrated suspensions and hamper the formation of gels (Figure \ref{fig:dfs40}-c): when this like-charged polymers coat the collapsed microgels, they both enhance electrostatic repulsion between the so-formed decorated objects and screen out the hydrophobic interactions originally driving the gel formation of bare microgels. 

With this in mind, we can also speculate, now more pertinently, that the (reentrant) gelation observed at $\xi=0.645$, can be ascribed to PE-mediated hydrophobic interactions, to the screening of the residual charge due to the increased counterion concentration and, since the mobility of the complexes does not varies remarkably for $\xi>$0.21, to depletion interactions due to unadsorbed chains. 

The encountered phenomenology is \rv{therefore} at odds with the assumption that highly sulfonated PSS is strictly a non-adsorbing polymer in the presence of PNIPAm microgels. Such an assumption has been adopted in the past \cite{rasmusson_flocculation_2004,snowden_temperature-controlled_1992} to justify the observed flocculation of PNIPAm microgels induced by PSS for $T\lesssim T_c$. If this was the case, the presence of PSS chains could only give rise to an enhanced depletion mechanism between the microgels driven by the unbalanced osmotic pressure exerted by the free chains. In concentrated suspensions such scenario would consequently result in a progressive gel strenghtening that is not observed, while by contrast the gel melting and the increase of the mobility modulus in dilute suspensions represent a robust indication that small PSS polymers adsorb on and partially interpenetrate hydrophobic microgels. In this respect, the presence of hydrophobic interaction between PNIPAm chains and non-sulfonated polystyrene is well documented in literature\cite{gao_coil--globule_1997}.  

We obtain a further insight on the charge restructuring process occurring in PE-microgel mixtures when we compare the mobility of the complexes to those of the pure PEs at the same concentration $C_{PE}$, namely the highest $\xi$ in the mixtures, below and above $T_c^H$ (Figure \ref{fig:tvsmob}-d,e,f). Remarkably, at T=20 $^\circ$C the average mobility modulus is always lower than the one measured in pure PE suspensions, signaling the absence of a large charge accumulation with respect to pure PEs. By contrast at 40 $^\circ$C, the mobility is always algebraically higher with respect to the pure PE solutions for the M-PLL and M-PDADMAC mxtures and it is comparable to that of the free PSS chains for M-PSS systems. The contrasting behavior observed below and above $T_c^H$ for cationic PEs confirms the occurrence of a net increase of the positive charge density in the presence of microgels for $T>T_c^H$. The latter can be due to both a compaction of the polyions and/or to an increase of the fraction of free counterions promoted by PE adsorption \cite{netz_complexation_1999}. We remind also, that a simple local increase of the polyion concentration would give rise to a decrease in mobility due essentially to an increased amount of condensed counterions in line with the data shown in figure \ref{fig:PEmob}-a,b and other results discussed in literature \cite{truzzolillo_counterion_2009}.
In the case of PSS chains this is less evident despite the mobility of the complexes is much larger in modulus compared to that measured for the pure PNIPAm microgels (Figure \ref{fig:tvsmob}-b).
A summary of the mobility values is also reported for comparison in Supplementary Material (Table 1).
\begin{figure*}[htbp]
 \centering
 \includegraphics[width=14cm]{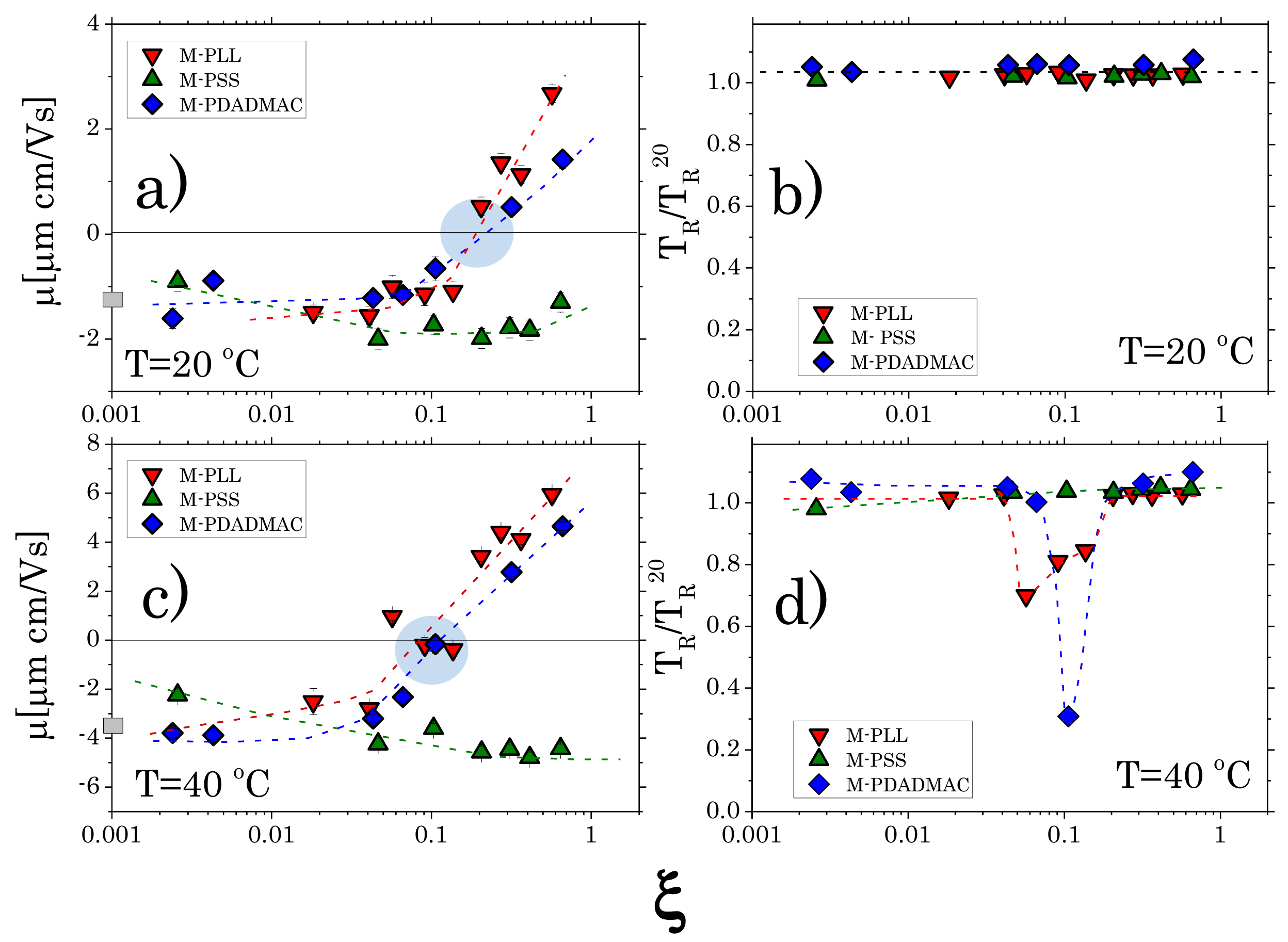}
 \caption{Mobility (a,c) and transmittance (b,d) of microgel suspensions ($\varphi=0.006$) at varying concentrations of PEs at $20^{\circ}C$ (a,b) and $40^{\circ}C$ (c,d). Error bars in a) and c) are the full widths at half maximum of mobility distributions. The blue circles in the panels (a,c) mark the isoelectric condition $\mu=0$.}
 \label{fig:MobTrans}
\end{figure*}
The charging and neutralization process of the complexes, and their stability at the two selected temperatures, can be finally rationalized if mobility and transmittance data are combined. This is shown in Figure \ref{fig:MobTrans}. For both M-PLL and M-PDADMAC mixtures a mobility reversal occurs at both 20 $^\circ$C and 40 $^\circ$C, while M-PSS mixtures show a progressive average mobility that becomes more and more negative as the PE concentration is increased (Figure \ref{fig:MobTrans}-a,c). Interestingly for both cationic PEs the mixtures undergo a reentrant condensation signalled by the large decrease of the relative transmittance (Figure \ref{fig:MobTrans}-d) in conjunction with the charge reversal observed at 40 $^\circ$C, while the suspensions stays stable at 20 $^\circ$C (Figure \ref{fig:MobTrans}-b). The simultaneous occurrence of an isoelectric condition and a large collodal condensation further confirm that the measured mobility has to be assigned to the PE-microgel complexes, as also pointed out by the sistematic unimodal mobility distribution measured for all the mixtures at any $T$ and by the sigmoidal temperature dependence of the mobility in the presence of PEs. 

\begin{figure}[htbp]
 \centering
 \includegraphics[width=8cm]{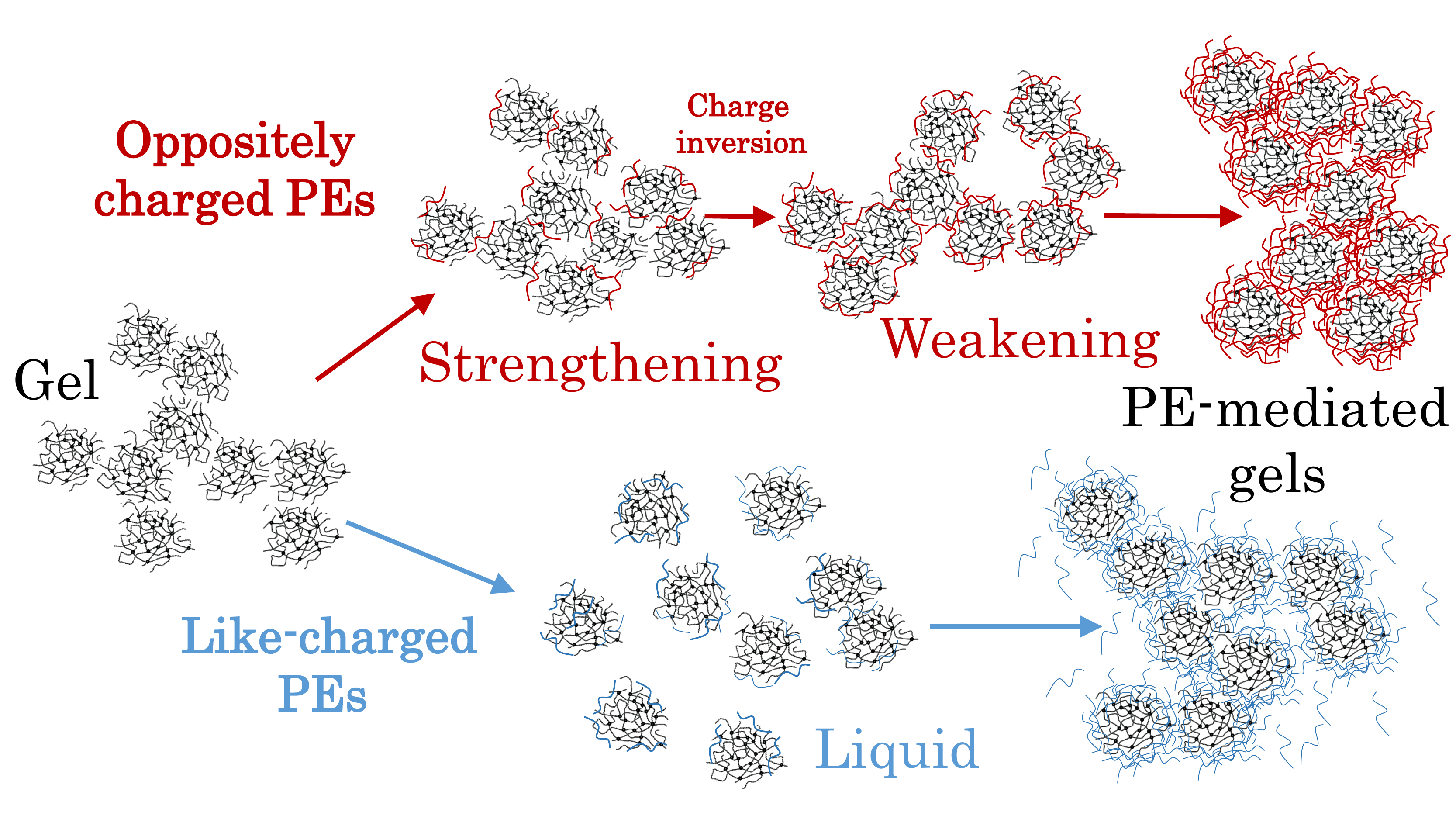}
 \caption{Schematic representation of the different states encountered when PEs are progressively added in suspensions of PNIAPm microgels and temperature is raised up to $T>T^H_c$.}
 \label{fig:sketch}
\end{figure}

Our data therefore confirm that the hydrophilic character of PNIPAm microgels below their VPT ($T<T_c^H$) hampers microgel condensation though an apparent isolectric condition is attained and chains partially adsorb onto microgels. By contrast above VPT ($T>T_c^H$) both hydrophobic interaction and a stronger charge-patch attraction between decorated microgels drive cluster formation localized at the isoelectric point, where the average mobility vanishes. We recall that charge patch attraction between decorated  colloids is enhanced when large surface charge density fluctuations characterize the colloidal surfaces\cite{velegol_analytical_2001,truzzolillo_kinetic_2009}. This is indeed the case for collapsed microgels, where bare portion of the microgel surfaces are supposedly densely charged and with an enhanced hydrophobicity.     
Therefore, when the destabilization due to polyelectrolyte adsorption is considered, PNIPAm microgels behave as charged colloids at high temperatures ($T>T_c^H$) and as nearly neutral colloids below their volume phase transition temperature ($T<T_c^H$), confirming previous results obtained in dilute suspensions of M-PLL mixtures\cite{sennato_double-faced_2021}. 
We have now shown that this has important repercussions on the rheology of concentrated dispersions whose viscoelasticity at high temperatures is largely affected by the presence of an isoelectric point. The emerging general scenario for $T>T_c^H$ is sketched in Figure \ref{fig:sketch}.   

Three final comments are in order.\\
i) The peak of the gel modulus (Figure \ref{fig:dfs40}-b) observed for mixtures of microgels and cationic PEs occurs at $\xi\simeq 0.04-0.05$ and it is slightly shifted to lower $\xi$ with respect to $\xi\simeq 0.1$, where the isoelectric condition and the reentrant condensation is observed in diluted samples (Figure \ref{fig:MobTrans}-c,d). This can be ascribed to the smaller volume accessible to PEs in concentrated microgel suspensions and therefore to a more effective neutralizing effect of the chains, that presumably spend much less time in their desorbed state. In addition to that, the counterion concentration is also larger in concentrated samples and this could also reduce the amount of oppositely charged polymers needed to neutralize the microgels, since the fraction of counterions condensed on the microgel is higher.\\    
ii) We note that at highest $\xi$ in dilute suspensions (Figure \ref{fig:MobTrans}) there is not evidence for large cluster formation, that would result in a decrease of the relative transmittance, while gel strengthening is reported for concentrated M-PSS and M-PLL mixtures. This again corroborates the hypothesis that these gels are networks stabilized by non-electrostatic interactions between the concentrated PE chains surrounding the microgels at high $\varphi$. PE concentration is much lower in the diluted suspensions used to measure the mobility and the transmittance of the PE-microgel complexes (Figure \ref{fig:MobTrans}). Here, the free volume accessible to PEs is much larger and their mutual contact is very much reduced.\\
iii) The charge inversion and re-entrant condensation occur far ($\xi\simeq 0.1$) from the nominal isoelectric condition ($\xi=1$). One possible explanation would be that most of the initiator ($\approx90\%$) does not participate to the polymerization, that then would result in a large residual mass present in the supernatant after the first post-synthesis centrifugation cycle. This is ruled out: after drying and removing the solvent from the supernatant we obtain a solid content equal to only $\approx1\%$ of
the total NIPAm monomers dissolved in water during the synthesis  and that comprises also the removed SDS. A  much larger solid content, containing both unreacted NIPAm monomers and initiator, would be present if the $\approx90\%$ of the initiator had not participated to the microgel formation.  
A second possibility is represented by the partial involvement of the ionic initiator in the adsorption process: within this scenario the very peripheral sulfonic groups anchored to the microgels are mostly free from condensed counterions and obviously more available for steric reasons to accomodate PE chains, while groups located further inside the microgel volume are on average more screened by confined counterions and they do not participate to the charge balance. This hypothesis is corroborated by recent experiments and simulations \cite{elancheliyan_role_2022} where the addition of randomly distributed charges barely affects the $R_g$ transition of PNIPAm-based microgels. Such a scenario, although it gives a plausible qualitative explanation for the discrepancy between the nominal and the observed isolectric condition, is certainly not a conclusive one, and the mechanisms of charge neutralization of microgels through polyion adsorption remain to be understood in the future.       

\section{Conclusions}\label{conclusion}
We have shown that adding polyelectrolytes bearing different charge has an important impact on the rheology of concentrated suspensions of anionic PNIPAm microgels, and that electrostatics influences crucially gel formation at high temperatures. We have studied the effect of the PEs on both a jammed glassy system, where PEs are confined in the interstitial spaces between faceted microgels or slightly penetrate into their peripheral corona, and when the concentrated suspension is brought beyond the microgel volume phase transition ($T>T_c^H$). We have shown that while jammed glasses are only weakly softened by PE addition, the impact of the PE chains is important at higher temperature, where PNIAPm microgels are charged hydrophobic colloids with high charge density. We find that a large and reentrant gel strengthening occurs in the presence of cationic PEs around the isoelectric point, where the average measured mobility is zero and large clustering takes place in diluted suspensions. Strikingly, for PE chains with residual hydrophobicity and bearing charges with the same sign as that of the microgels, we observe a melting of the original gel phase at $T>T_c^H$. Mobility measurements suggested that also in this case there is at least a partial adsorption and/or penetration of the PEs into the microgels, increasing their net charge, screening the hydrophobic attraction between collapsed microgels and hence hampering the gelation of the system within the probed experimental time scale. \rv{Non-electrostatic interactions are thus also very important, especially when microgels are in their collapsed state, and at high PE concentrations, far from the isoelectric point, we also find that the rheology of the mixtures is highly influenced by these specific effects}: for mixtures containing partially hydrophobic PLL or PSS chains, gelation is favoured by the presence of a high PE content, while for the more hydrophilic PDADMAC polymers this is not observed. Our work shows that the rheology of thermosensitive ionic microgels can be tuned by adsorbing polyelectrolytes, and confirms that PNIAPm microgels must be considered as highly charged hydrophobic colloids at high temperatures. Our study paves the way for more systematic investigations including PEs sharing the same chemistry but with different molecular weights and degree of hydrophobicity (e.g PSS chains with different $M_w$ and sulfonation degrees) that are required to discern more in depth the role played by the chain size and the non-electrostatic effects in these composite soft materials. \rv{In this respect, the use of fluorescent PEs and molecular dynamic simulations, would surely help to unravel the role played by residual unadsorbed chains on the overall microgel stability.} Finally, whether PE adsorption causes glass melting at $T<T_c^H$ or reentrant gelation at $T>T_c^H$  and lower generalized volume fractions is still unknown and deserves a thorough investigation.   


\section*{Conflicts of interest}
There are no conflicts to declare.

\section*{Acknowledgements}
We acknowledge financial support from the Agence Nationale
de la Recherche (Grant ANR-20-CE06-0030-01; THELECTRA). D.T. thanks Dr. S. Sennato for fruitful discussions.   



\balance

\renewcommand\refname{References}

\bibliography{rsc} 
\bibliographystyle{rsc} 

\end{document}


\section{Light scattering}
\begin{figure}[ht]
 \centering
    \includegraphics[width=18cm]{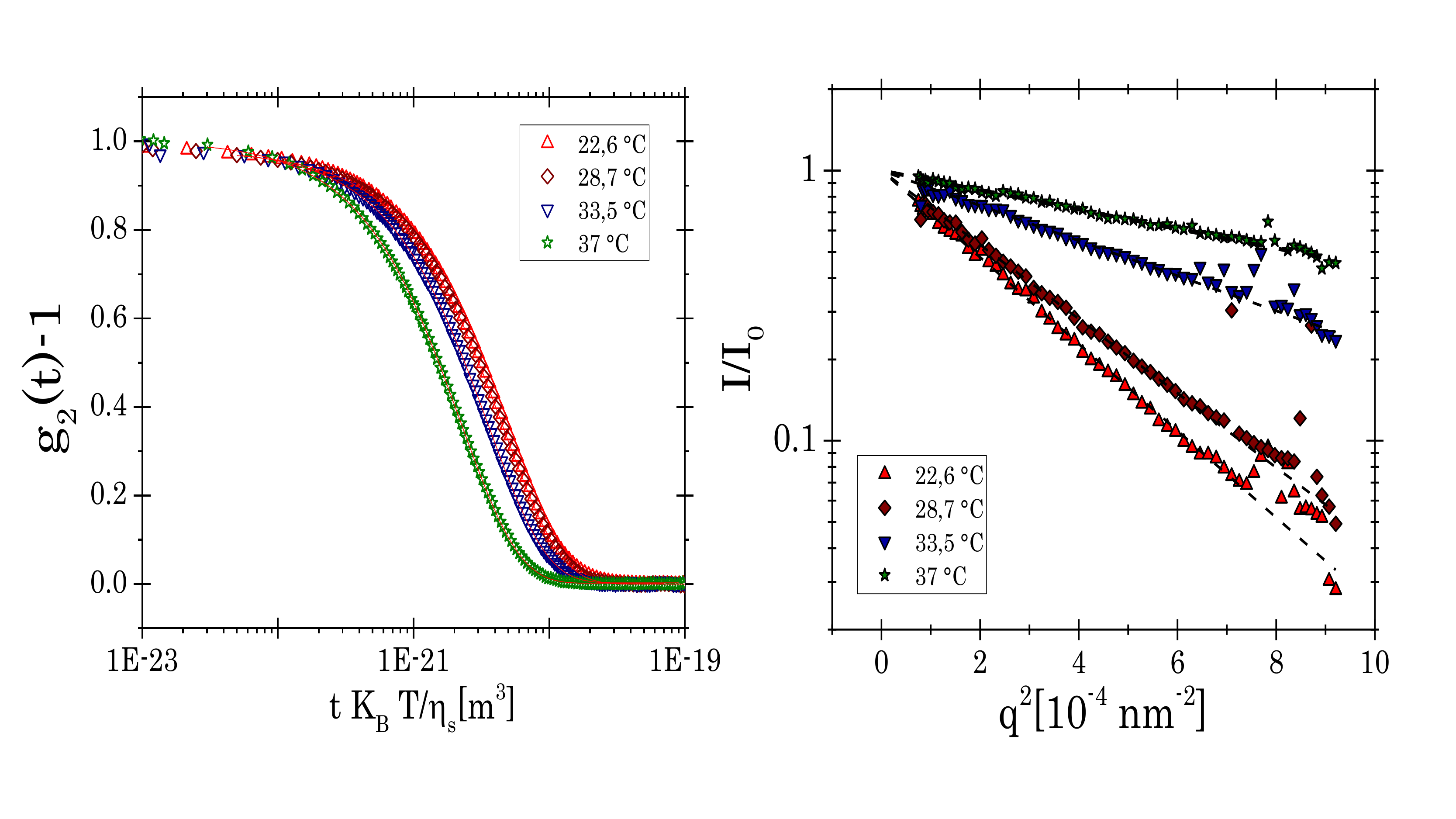}
    \caption{Autocorrelation functions $g_2-1$ (left panel) and normalized scattered intensity $I/I_0$ vs $q^2$ (right panel) at different temperatures as indicated in the panels. Red solid (black dashed) lines are cumulant (Guinier) fits according to equations 3 and 4 of the main manuscript for $g_2(t)-1$ and $I/I_0$ respectively.}
    \label{fig:DLSSLS}
\end{figure}

We report in figure \ref{fig:DLSSLS} selected intensity autocorrelation functions ($g_2(t)-1$) and time-averaged scattered intensity $I/I_0$ as a function of the squared modulus of the scattering vector $q^2$ at different temperatures. The lag time $t$ (left panel) is normalized by $\eta_s/K_BT$ to filter out the trivial speeding up of the dynamics due to the temperature dependence of the solvent
viscosity $\eta_s$ and the thermal energy $k_BT$. $I_0$ is obtained by fitting $I(q)$ to the Guinier equation (eq. 4 of the main text).
The autocorrelation functions and the scattered intensity data have been fitted by using respectively equation 3 and 4 (section 2.5 of the main manuscript) to extract the hydrodynamic radius ($R_H$) and gyration radius ($R_g$) as a function of temperature.

\section{Viscosimetry}

\begin{figure}[htbp]
 \centering
    \includegraphics[width=12cm]{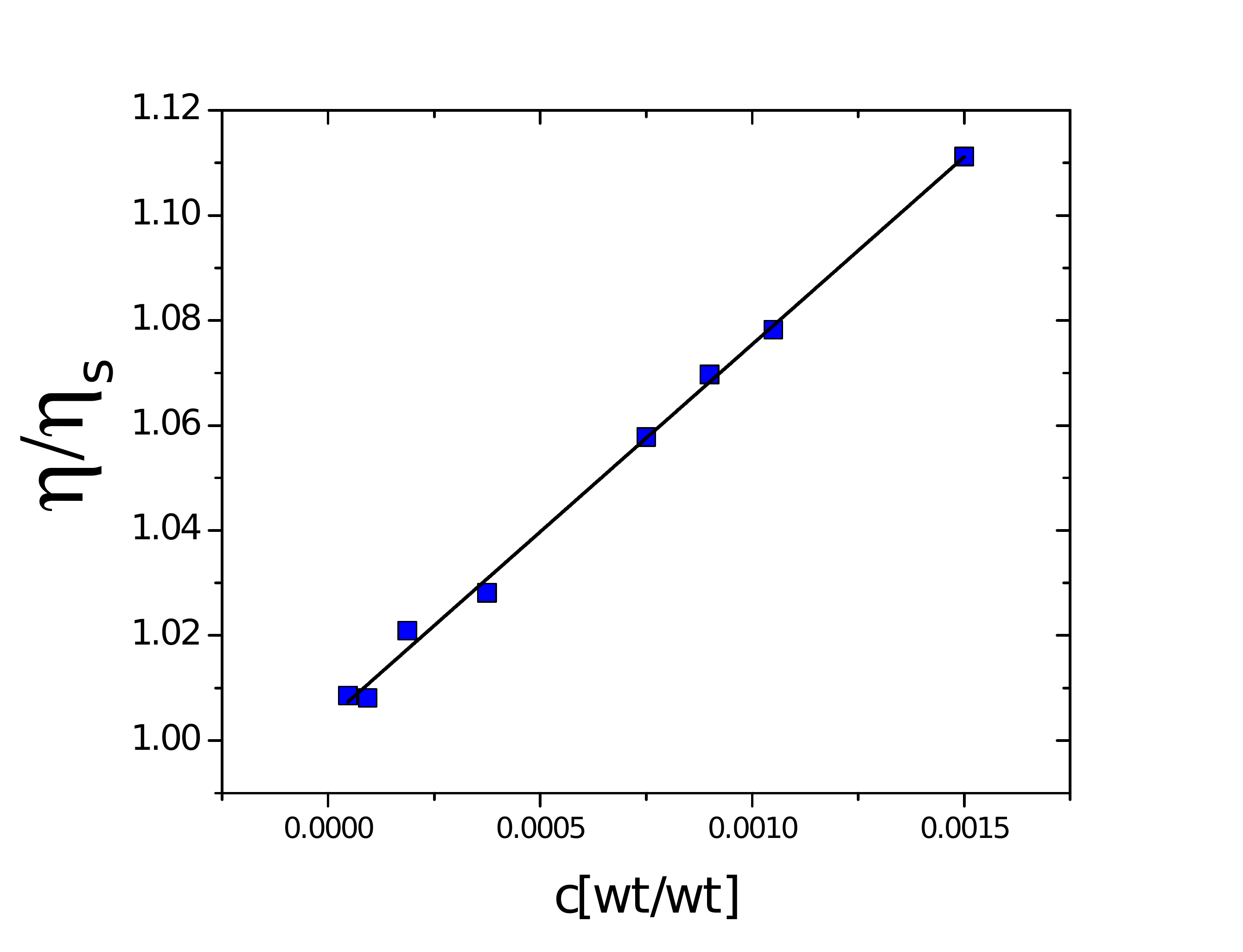}
    \caption{Viscosity ratio $\eta/\eta_s$ for very dilute suspensions of microgels. The straight line is a linear fit of the data to the Einstein equation.}
    \label{fig:ETA}
\end{figure}

Figure \ref{fig:ETA} shows the viscosity ratio $\eta/\eta_s$ for very dilute suspensions of bare microgels in the range 4.68 $\cdot$ 10$^{-5}$ $<$ $c$(wt/wt) $<$  1.5 $\cdot$ 10$^{-3}$. The proportionality constant $k$ between the generalized volume fraction $\varphi$ and the mass fraction $c$ of microgels has been obtained, as detailed in the main text, via a linear fit of the data to the Einstein equation.

\section{\rv{Rheology of bare microgel suspensions}}
\begin{figure}[htbp]
\centering
  \includegraphics[width=12cm]{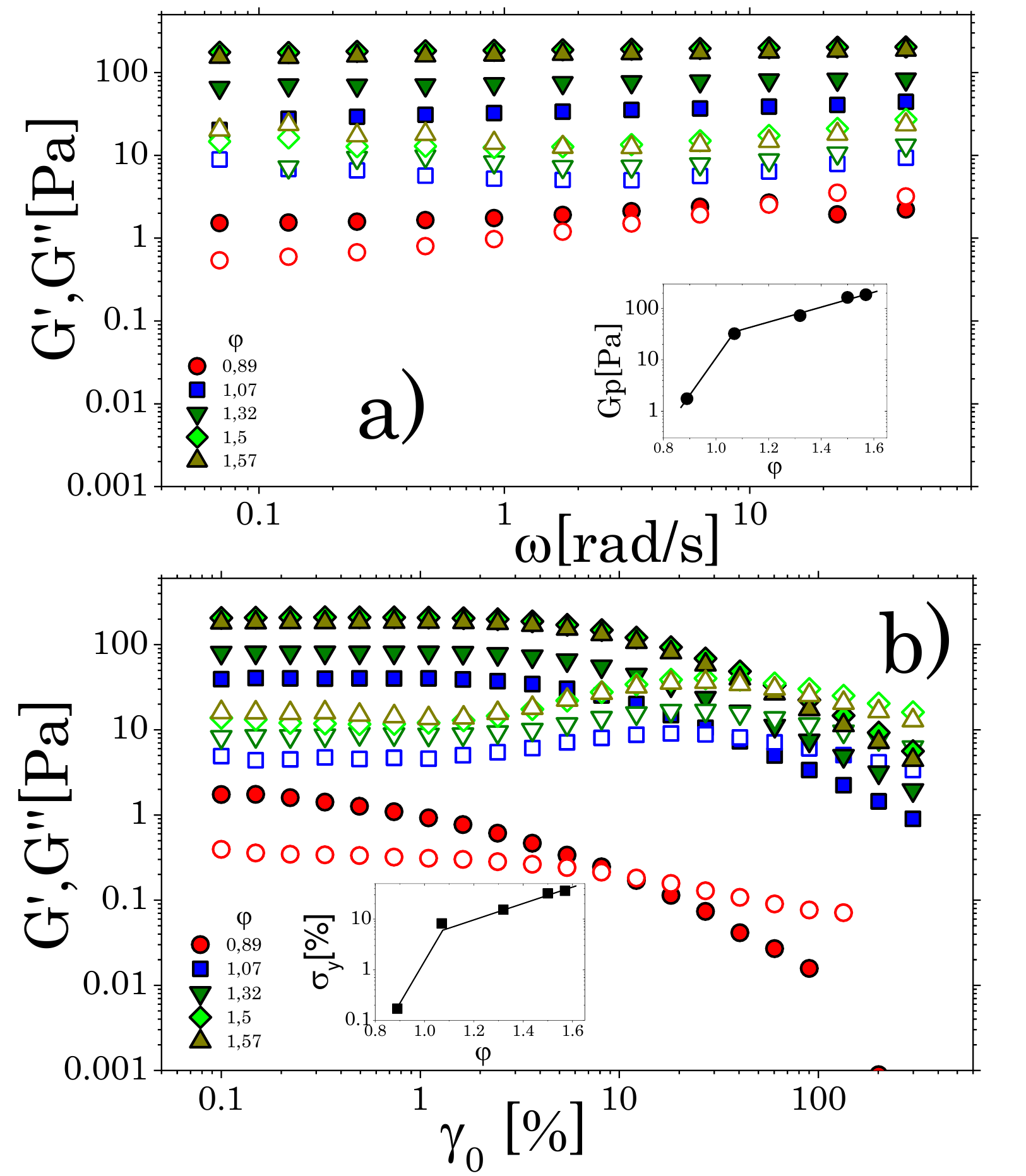}
  \caption{\rv{Storage modulus G' (solid symbols) and loss modulus G" (open symbols) as a function of the oscillatory frequency $\omega$ (a) and respective first harmonic moduli as a function of the strain amplitude $\gamma_0$ at $\omega=1$ rad/s (b) of concentrated PniPAm microgel suspensions at T=20 $^\circ$C and different volume fractions as indicated in the panels. The insets show the plateau modulus $G_p$ (a) and the yield stress $\sigma_y$ (b) as a function of $\varphi$.}}
  \label{fig:baremicrogelsDFSDSS}
\end{figure}

\rv{The rheology of bare microgels suspensions was investigated as a function of $\varphi$ at T=20 $^\circ$C to determine precisely the rheological state of the suspension in which PEs are progressively added. Figure \ref{fig:baremicrogelsDFSDSS} shows dynamic frequency (a) and strain (b) sweep tests performed in the range 0.89 $\leq\varphi\leq$ 1.57. All samples show a solid-like response and a typical glassy behavior for $\varphi\geq$ 1.07, with $G'>G"$ over about 3 decades in frequency and shallow minimum of $G"$ \cite{sessoms_multiple_2009,mewis_colloidal_2011}. Between  $\varphi$=1.07 and $\varphi$=0.89 we observe a large drop of both the moduli, which flags the proximity to the rheological glass transition. 
The DSS tests confirm such a scenario with all the suspensions showing a nearly strain-independent first-harmonic moduli at low strains before non-linear behavior appears, with $G"(\gamma_0)$ reaching a maximum and then declining for $\varphi\geq$ 1.07. The maximum of $G"(\gamma_0)$ is absent for $\varphi$=0.89 pointing to an important reduction of dissipative processes involved in the yielding transition. In all the cases a crossover between the two first harmonic moduli occurs and samples yield under oscillatory strain.     
The two insets show the storage modulus $G_p$=$G'$($\omega$=1 rad/s) (a) and the yield stress $\sigma_y$ (b) at which the crossover $G'(\gamma_0)=G"(\gamma_0)$ occurs as a function of $\varphi$. The two quantities show an affine behavior with a drop of their values occurring between $\varphi$=1.07 and $\varphi$=0.89. Pellet and Cloitre\cite{pellet_glass_2016} have attributed such a sharp change to the passage from a thermal to a jammed glass of microgels, with the former being a solid whose elasticity is dominated by entropic caging, while the latter is ruled by contact forces between particles and particle deformability. Therefore the microgel concentration that characterize the PE-microgel mixtures ($\varphi=1.5$) is a jammed glass at T=20 $^{\circ}$C. We show in the main manuscript (section 3.2) that at this number density, microgels form a gel when the suspension is heated up to T=40 $^\circ$C.
}

\section{Ageing of bare microgel suspensions}
\begin{figure}[ht]
	\centering
	\includegraphics[width=14cm]{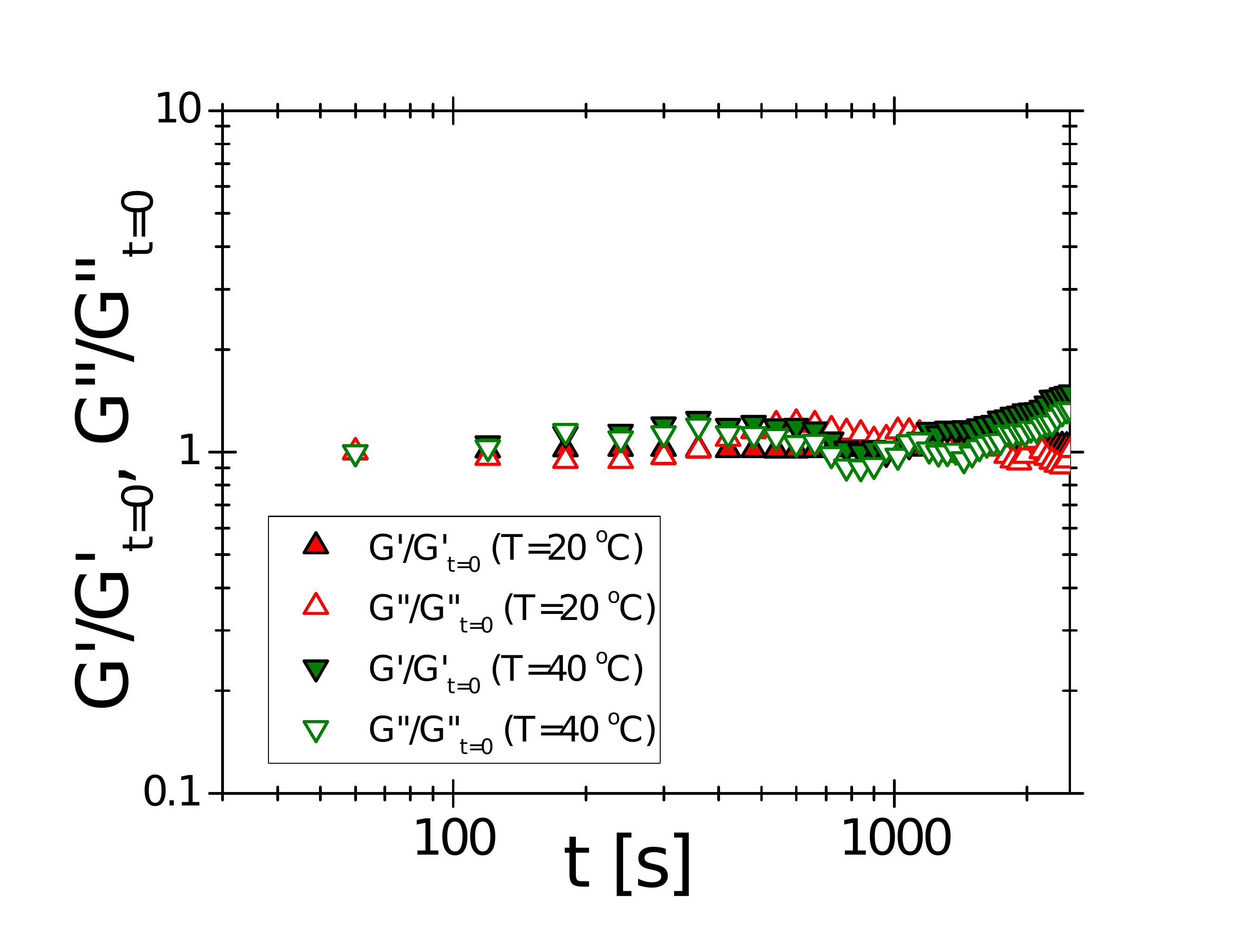}
	\caption{Normalized viscoelastic moduli ($\omega=$2 rad/s) of the bare microgel suspension ($\varphi=1.5$, $\xi=0$) at T=20$^{\circ}$C and T=40$^{\circ}$C as indicated in the panels.}
    \label{fig:DTS}
\end{figure}

Prior to mixture preparation we performed a dynamic time sweep experiment to evaluate the aging of the bare concentrated microgel suspension ($\varphi=1.5$, $\xi=0$) over a time approximately equal to the duration of one whole experiment (t$\approx$ 2500 s). Figure \ref{fig:DTS} shows the normalized storage and loss modulus ($G'/G'_{t=0}$, $G"/G"_{t=0}$) of the sample at $\omega=$2 rad/s. The moduli did not show any remarkable evolution within this time window. This excludes that the changes of the moduli observed for the mixtures are due to different age of the pure microgel system.   

\section{Mobility}
\begin{figure}[htbp]
	\centering
	\includegraphics[scale=0.5]{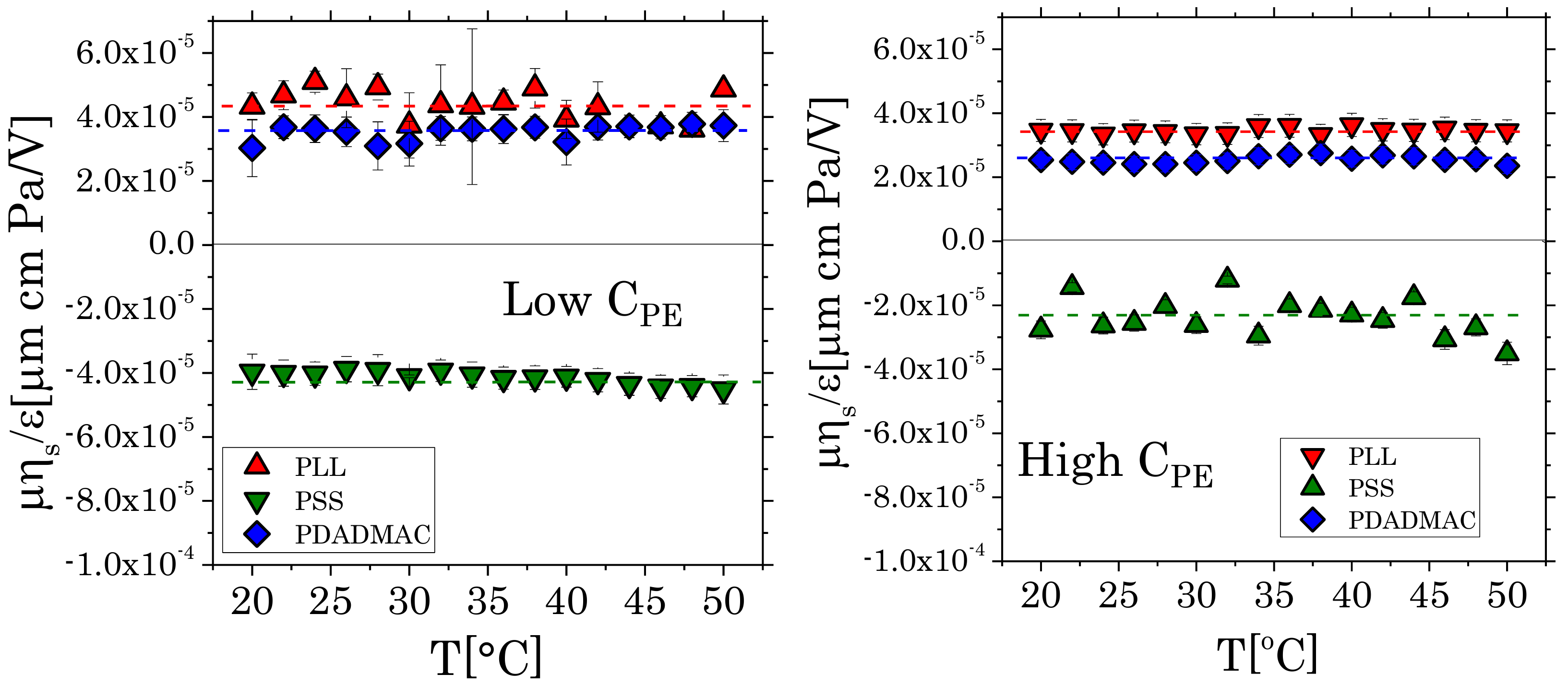}
	\caption{Normalized electrophoretic mobility of PLL at  $C_{PE}$=1.25 mg/ml (left) and $C_{PE}$=28 mg/ml (right),  PSS at $C_{PE}$=1.75 (left) and $C_{PE}$=5.1 (right), PDADMAC at $C_{PE}$=1.38 mg/ml (left) and $C_{PE}$=35 mg/ml (right) as a function of temperature.}
    \label{fig:mobnorm-PE}
\end{figure}

\begin{figure}[htbp]
	\centering
	\includegraphics[scale=0.5]{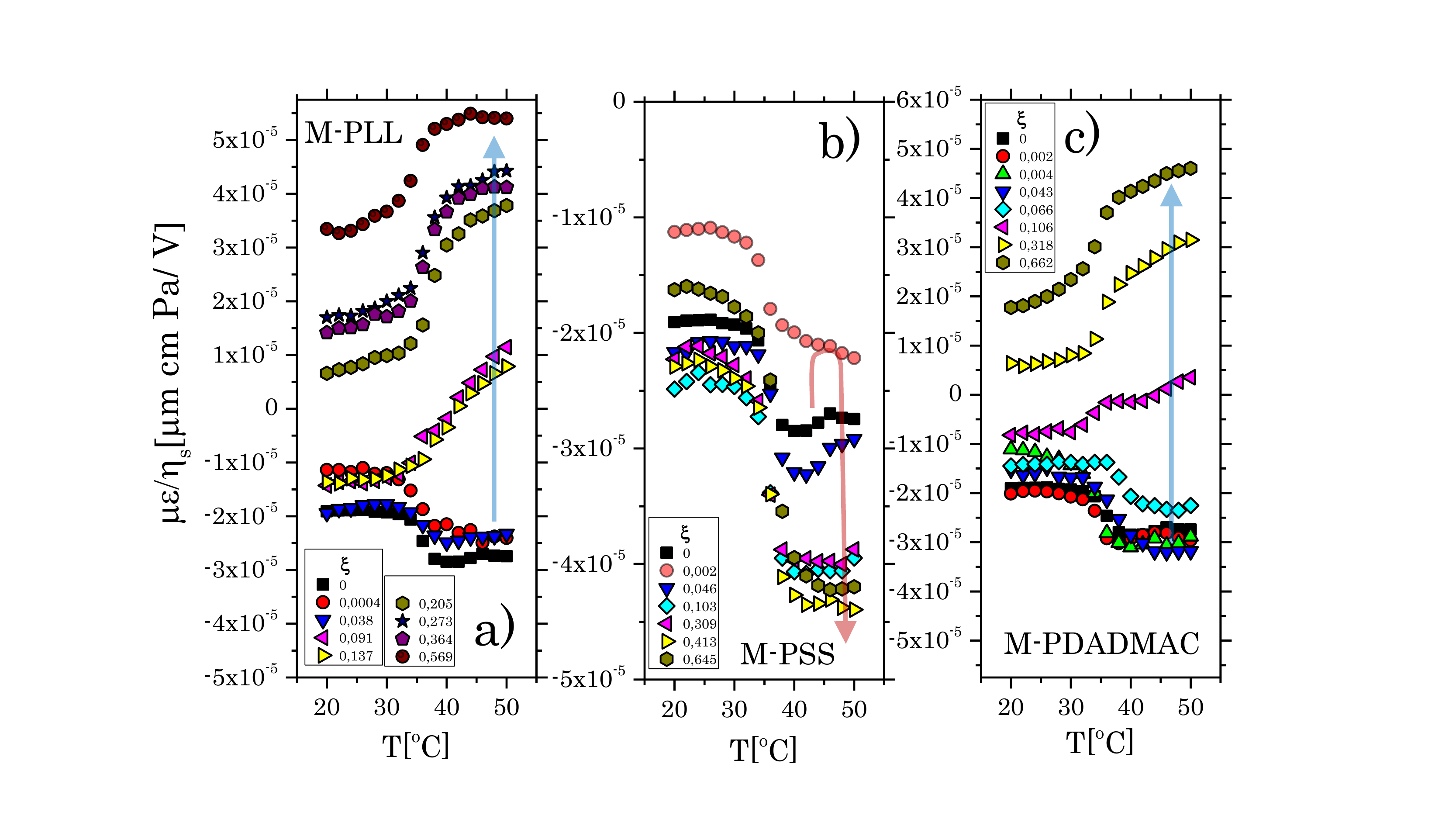}
	\caption{Normalized electrophoretic mobility of a) M-PLL, b)  M-PSS and c) M-PDADMAC complexes at varying PE concentrations as a function of temperature. Data contains complementary samples that were not characterized via rheology. The colored arrows are a guide for the eye and point to the mobility variation from $\xi=$0 up to the maximum PE concentration at $T>T_c^H$.}
    \label{fig:MOBILITY}
\end{figure}

We report below the normalized mobility $\mu\eta/\epsilon$ for both the pure PE suspensions (Figure \ref{fig:mobnorm-PE}) and the PE-microgel mixtures (Figure \ref{fig:MOBILITY}). The data are the same shown in figures \rv{4 and 7} of the main manuscript. 
For pure PE suspensions the normalized mobility does not show a visible trend for increasing temperature, indicating that the observed variation in absolute mobility is due mainly to the reduction of the solvent viscosity going 20 $^{\circ}$C to 50 $^{\circ}$C. We recall that in this temperature range the viscosity of water $\eta_s$ decreases by a factor $\approx$1.85, while the relative permittivity $\varepsilon$ decreases by a factor $\approx$ 0.86. By contrast the PE-microgel mixtures maintain the trend showed by the mobility $\mu$ (figure \rv{7} of the main text), hence pointing to an important change of the charge density of the complexes upon varying temperature.    

\begin{table}[H]
\caption{Mobility in $\mu$m cm V/s of PEs solutions and PE-microgel mixtures at T=20 $^{\circ}$C and T=40 $^{\circ}$C for the highest PE concentrations. Data are shown in figure 4 and figure \rv{7} of the main manuscript. Data for pure microgel suspensions are also given for reference.}\label{tablesynth}
\centering
\begin{tabular}{ccc}
\toprule
\multicolumn{1}{l}{\textbf{{\footnotesize Sample code}}}	& \textbf{{\footnotesize $\mu$(T=20 $^{\circ}$C)}}	& \textbf{{\footnotesize $\mu$(T=40 $^{\circ}$C)}}\\
\midrule
\multicolumn{1}{l}{M (No PE)}		& -1.52 $\pm$0.10			& -3.20$\pm$ 0.16\\
\multicolumn{1}{l}{PLL ($C_{PE}=$1,25 mg/ml)}		& 3.46 $\pm$0.28			& 4.40$\pm$ 0.66\\
\multicolumn{1}{l}{M-PLL ($C_{PE}=$1,25 mg/ml, $\xi=$0.5691)}		& 2.67$\pm$	0.16		& 5.94$\pm$0.19\\
\multicolumn{1}{l}{PSS ($C_{PE}=$ 1.75 mg/ml)}		& -3.17$\pm$0.44			& -4.62$\pm$0.36\\
\multicolumn{1}{l}{M-PSS ($C_{PE}=$ 1.75 mg/ml, $\xi=$0.645)}		& -1.30$\pm$0.18			& -4.43$\pm$0.19\\
\multicolumn{1}{l}{PDADMAC ($C_{PE}=$ 1.38 mg/ml)}		& 2.42$\pm$0.71			& 3.60$\pm$0.40\\
\multicolumn{1}{l}{M-PDADMAC ($C_{PE}=$ 1.38 mg/ml, $\xi=$0.6625)}		& 1.42$\pm$0.12			& 4.64$\pm$0.18\\
\bottomrule
\end{tabular}
\end{table}

\newpage
\bibliography{rsc}